\documentclass[epsfig,12pt]{article}
\usepackage{epsfig}
\usepackage{graphicx}
\usepackage{array}
\usepackage{color}
\usepackage{bm}

\newcommand{\beq}{\begin{equation}}   
\newcommand{\eeq}{\end{equation}}
\newcommand{\beqn}{\begin{eqnarray}}   
\newcommand{\eeqn}{\end{eqnarray}}

\def\ntwot{${\mathcal N}=(2,2)\;$}
\def\ntwoo{${\mathcal N}=(0,2)\;$}

\newcommand{\gsim}{\lower.7ex\hbox{$
\;\stackrel{\textstyle>}{\sim}\;$}}
\newcommand{\lsim}{\lower.7ex\hbox{$
\;\stackrel{\textstyle<}{\sim}\;$}}
\begin{document}

\begin{titlepage}

\begin{flushright}
FTPI-MINN-14/1, UMN-TH-3321/14\\
January 26\\
\end{flushright}

\vspace{0.3cm}

\begin{center}
{  \large \bf  Two-Dimensional Sigma Models Related to\\[2mm]
Non-Abelian Strings in Super-Yang-Mills\footnote{Invited paper, to be published in the Pomeranchuk Memorial Volume (2014).}}

\vspace{6mm}

{\em
Dedicated to the 100$^{\,th}$ birthday of Isaak Yakovlevich Pomeranchuk}

\end{center}
\vspace{0.3cm}

\begin{center}
 {\large 
 M. Shifman$^a$ and A. Yung$^{a,b}$}
\end {center}

 
\begin{center}

$^a${\em William I. Fine Theoretical Physics Institute, University of Minnesota,
Minneapolis, MN 55455, USA}

\vspace{2mm}

$^{b}${\em Petersburg Nuclear Physics Institute, Gatchina, St. Petersburg
188300, Russia}

\end {center}

\vspace{0.5cm}

\begin{center}
{\large\bf Abstract}
\end{center}

We review diverse two-dimensional models  emerging on the world sheet of non-Abelian strings in the low-energy limit. Non-Abelian strings are supported in a class of four-dimensional bulk theories with or without supersymmetry.
In supersymmetric bulk theories we are mostly interested in BPS-saturated strings.
Some of these two-dimensional models, in particular, heterotic models,  were 	
scarcely studied in the past, if at all. Our main emphasis is on the heterotic ${\mathcal N}=(0,2)$ models. 
We describe their large-$N$ solution.
We briefly comment on ${\mathcal N}=(0,1)$ models although so far they are not obtained on the the world sheet of non-Abelian strings.

\end{titlepage}

\newpage
{\small
\tableofcontents
}
\newpage

\rule{0mm}{9mm}

\vspace{3cm} 

\begin{center}
{\Huge Part I\\[6mm]

Introductory}
\end{center}

\vspace{3cm}

\begin{center}
\includegraphics[width=4in]{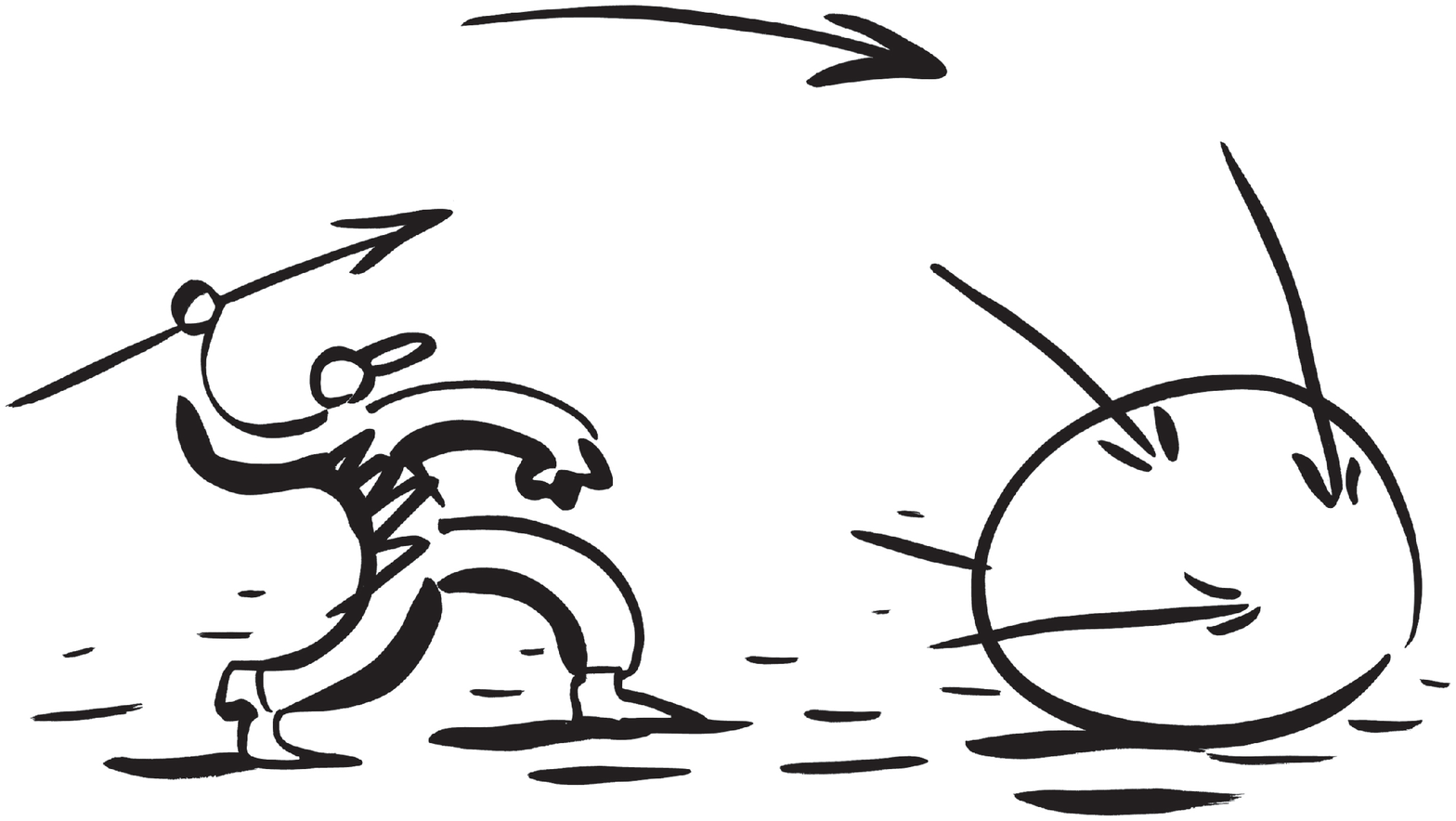}
\end{center}

\newpage

\section{Introduction}

Isaak Yakovlevich Pomeranchuk, the founder of the ITEP Theory Department, died in December of 1966,
only six years before the advent of revolutionary changes in high-energy physics. His work with Landau \cite{LP}
(see also \cite{P2})
on the so-called Moscow zero charge (currently known as infrared freedom in Abelian gauge theories),
shaped the subsequent research on gauge theories which culminated in 1973, with the discovery of asymptotic freedom in non-Abelian gauge theories \cite{GWP}. Non-Abelian gauge theories proved to be the basis of the modern theory.\footnote{Pomeranchuk witnessed the discovery of a two-dimensional asymptotically free field theory as early as in 1958 \cite{Ans}, but at that time due attention was not paid to this work.}

Asymptotic freedom is just one aspect of these theories. Another aspect is a unique behavior in the infrared domain, at strong coupling, known as  {\em confinement}, or, sometimes, color confinement. Despite four decades of vigorous efforts analytic understanding of the phenomenon of color confinement in quantum chromodynamics is still incomplete. At the same time significant advances occurred in 1994 when Seiberg and Witten solved ${\mathcal N}=2$ super-Yang-Mills theory \cite{SW}.

In the mid-1970s Nambu,   't Hooft, and  Mandelstam (independently) 
put forward an idea \cite{NTM}
of a ``dual Meissner effect" as the underlying mechanism for color confinement.
Within their conjecture, in appropriate Yang-Mills theories chromomagnetic   ``monopoles" condense
leading to formation of ``chromoelectric flux tubes" between the probe quarks.
At that time the  Nambu-'t Hooft-Mandelstam paradigm was not even a physical 
scenario, rather a
dream, since people had no clue as to
the main building blocks such as non-Abelian  
flux tubes.

The Seiberg-Witten solution \cite{SW} triumphantly demonstrated the emergence of the confining strings
as a result of a small ${\mathcal N}=1$ deformation ${\mathcal N}=2$ super-Yang-Mills theory.

However, although these strings appear in the non-Abelian theory they turned out to be Abelian in their structure \cite{SS},
in essence identical to the Abrikosov-Nielsen-Olesen (ANO) strings \cite{ANO}.

Just like the fundamental string in string theory, the ANO string (at low excitation energies) is fully
characterized by the position of its center in the perpendicular plane Ð the so-called translational moduli. 
The orientation of the magnetic flux in the string core is rigidly fixed in the SW solution. Say, for the SU(2) gauge group it can be aligned along the third axis in the color space. Shortly after the SW discovery it was realized
that for QCD-like theories, in which there are no preferred directions in the color space, it would be more appropriate to have the flux in the string core   fluctuating freely ``inside" the non-Abelian group.  In other words for QCD strings it is desirable to have additional orientational moduli on the string world sheet. Such strings became known 
as non-Abelian. 

The search for genuinely non-Abelian strings started in the end of 1990s and culminated in their discovery 
\cite{AHDT} in 2003. Dynamics of the extra -- orientational -- moduli on the string world sheet was 
demonstrated to be described by CP$(N-1)$ model, where $N$ is the number of colors in the bulk theory. 
Since then a large variety of non-Abelian strings became known; some of them 
support two-dimensional theories that
had been  known for decades, others exhibit nontrivial and largely unexplored sigma models on the string world sheet. This review is devoted to two-dimensional sigma models which came into the limelight
in connection with the non-Abelian strings. The review is by necessity brief and represents,  in a sense,
a travel guide in this subject.

Historically nontrivial sigma models on the string world sheet first emerged in the context of
supersymmetric bulk theories. Now it is clear that supersymmetry is not necessary, nonsupersymmetric bulk theories
can support them too \cite{odi,dva}. Due to the fact that we will mainly focus on least explored 
world-sheet theories -- heterotic two-dimensional sigma models -- our discussion will be tied up with supersymmetry.
A significant part of this review is devoted to results which we obtained after 2008.
For a review before 2009 see \cite{SYbook}.
\section{How world-sheet models appear: \\
the simplest example}
\label{sec2}

The simplest and historically the first model supporting non-Abelian strings is 
 \cite{{AHDT}}  ${\mathcal N}=2$ super-Yang--Mills theory with
  the number of colors equal to the number of flavors (i.e. if the gauge group is SU(2), to which we will limit ourselves in this section,  we introduce two (s)quark
  flavors). Moreover, we add a U(1) factor to the gauge group, so that, in fact, the gauge group is U(2). We endow this U(1) factor with the Fayet--Iliopoulos term $\xi$
  \cite{FI}. The latter is needed to make non-Abelian strings BPS-saturated. BPS saturation is {\em not} a necessary condition. However, it simplifies calculations.
  
 The bosonic part of the basic U(2)
theory with two flavors has the form \cite{{AHDT}} (in the Euclidean space)
\beqn
{\mathcal L}&=& \frac1{4g^2_2}
\left(F^{a}_{\mu\nu}\right)^2 +
\frac1{4g^2_1}\left(F_{\mu\nu}\right)^2
+
\frac1{g^2_2}\left|D_{\mu}a^a\right|^2 +\frac1{g^2_1}
\left|\partial_{\mu}a\right|^2  
\nonumber\\[4mm]
&+&  \left|\nabla_{\mu}
q^{A}\right|^2 + \left|\nabla_{\mu} \bar{\tilde{q}}^{A}\right|^2
+V(q^A,\tilde{q}_A,a^a,a)\,.
\label{qed}
\eeqn
Here $D_{\mu}$ is the covariant derivative in the adjoint representation
of  SU$(2)$, and
\beq
\nabla_\mu=\partial_\mu -\frac{i}{2}\; A_{\mu}
-i A^{a}_{\mu}\, T^a\,, \qquad T^a= \frac 12 \tau^a\,,
\label{defnabla}
\eeq
where $\tau^a$ are the Pauli matrices acting in the color SU(2) group.
The coupling constants $g_1$ and $g_2$
correspond to the U(1)  and  SU$(2)$  sectors, respectively.
With our conventions, the U(1) charges of the fundamental matter fields
are $\pm1/2$. Two squark fields are denoted by $q^A$ and $\tilde{q}_A$, respectively (the flavor index 
$A=1,2$).
The doubling of the (s)quark fields is required by ${\mathcal N}=2$ supersymmetry.
In addition to the flavor index $A$ the the (s)quark fields carry SU(2) doublet index too;
therefore, they can be viewed as a $2\times2$ matrix. Moreover, $a^c$ ($c=1,2,3$) is the complex
scalar field in the adjoint representation of SU(2), the superpartner of the SU(2) gauge bosons,
while $a$ without the superscript is the superpartner of the U(1) gauge boson. For brevity we will refer to these fields as to ``adjoints."

\vspace{1mm}

The potential $V(q^A,\tilde{q}_A,a^a,a)$ in the Lagrangian (\ref{qed})
is a sum of  $D$ and  $F$  terms,
\beqn
&& V(q^A,\tilde{q}_A,a^a,a)  = 
 \frac{g^2_2}{2}
\left( \frac{i}{g^2_2}\,  \varepsilon^{abc} \bar a^b a^c
 +
 \bar{q}_A\,T^a q^A -
\tilde{q}_A T^a\,\bar{\tilde{q}}^A\right)^2
\nonumber\\[3mm]
&&+ \frac{g^2_1}{8}
\left(\bar{q}_A q^A - \tilde{q}_A \bar{\tilde{q}}^A \right)^2
+ 2g^2_2\left| \tilde{q}_A T^a q^A \right|^2+
\frac{g^2_1}{2}\left| \tilde{q}_A q^A -\xi \right|^2
\nonumber\\[3mm]
&&+\frac12\sum_{A=1}^N \left\{ \left|(a+\sqrt{2}m_A +2T^a a^a)q^A
\right|^2
+
\left|(a+\sqrt{2}m_A +2T^a a^a)\bar{\tilde{q}}^A
\right|^2 \right\}.
\label{pot}
\eeqn
Here   $m_A$ are the (s)quark mass terms, and 
the sum over the repeated flavor indices $A=1,2$ is implied.

Let us discuss the vacuum structure of this model. 
Nonvanishing of the Fayet-Iliopoulos term $\xi\neq 0$ implies an isolated vacuum  with the maximal possible value
of condensed (s)quarks -- two. 
The  vacua of the theory (\ref{qed}) are determined by the zeros of 
the potential (\ref{pot}). The adjoint fields develop the following vacuum expectation values (VEVs):
\beq
\langle \Phi\rangle = - \frac1{\sqrt{2}}
 \left(
\begin{array}{cc}
m_1 &  0 \\
0 & m_2\\
\end{array}
\right),
\label{avev}
\eeq
where we defined the scalar adjoint matrix as
\beq
\Phi \equiv \frac12\, a + T^a\, a^a.
\label{Phidef}
\eeq
If
$
m_1=m_2
$ and $\xi =0$
 the  SU$(2)\times$U(1)
gauge group remains classically unbroken, 
since in this case $m$ can be absorbed in $a$. Alternatively, we can set  $m=0$ from the beginning.
However, if $m_1\neq m_2$ SU(2) is broken down to U(1). Furthermore, if $\xi\neq 0$
we must take into account the squark VEVs which results in  Higgsing of all gauge bosons..

In the vacuum the squark VEVs have a peculiar color-flavor locked  form
\beqn
\langle q^{kA}\rangle &=&\langle \bar{\tilde{q}}^{kA}\rangle =\sqrt{
\frac{\xi}{2}}\,
\left(
\begin{array}{cc}
1 &  0 \\
0 &  1\\
\end{array}
\right),
\qquad
k=1,2,\qquad A=1,2\, .
\label{qvev}
\eeqn
The potential (\ref{pot}) vanishes if $\Phi $ and $q$ are chosen according to (\ref{avev}) and (\ref{qvev}), respectively.
($\xi$ is assumed to be large, $\xi\gg\Lambda^2$, to warrant quasi classical treatment.)

The  vacuum field (\ref{qvev}) results in  the spontaneous
breaking of both gauge and flavor SU($2$)'s.
A diagonal global SU($2$) survives, however,
\beq
{\rm U}(2)_{\rm gauge}\times {\rm SU}(2)_{\rm flavor}
\to {\rm SU}(2)_{C+F}\,.
\label{c+f}
\eeq
Thus, a color-flavor locking takes place in the vacuum.

\vspace{1mm}

Why does the model described above support a novel type of strings, non-Abelian?

\vspace{1mm}

\begin{figure}[h]
\epsfxsize=5cm
\centerline{\epsfbox{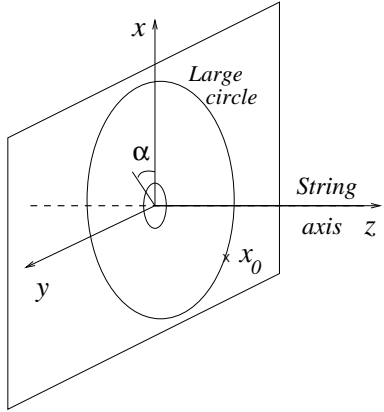}}
\caption{\small
Geometry of the string.}
\label{mmmon}
\end{figure}

The conventional ANO string corresponds to a U(1) winding of the phase of all squark fields in the plane, perpendicular to the
string axis,
\beq
q^{kA}\longrightarrow\sqrt{
\frac{\xi}{2}}\,e^{i\alpha}
\left(
\begin{array}{cc}
1 &  0 \\
0 &  1\\
\end{array}
\right),
\eeq
where $\alpha$ is the polar angle in the perpendicular plane (see Fig. \ref{mmmon}). Its topological stability is due to $\pi_1({\rm U}(1)) = Z$.
Now we have more options, however, due to the fact that $ {\rm SU}(2)_{C+F}$ has center. Usually people say that $\pi_1({\rm SU}(2)) $ is trivial and, therefore there are no other topologically stable strings.

This is not quite the case in the model at hand. Observe that the center of the SU(2) group, $Z_2$, belongs to the U(1) factor too. This means
that we can split the $2\pi$ windings in two halves: the first (from 1 to $-1$) is carried out in U(1), while the second,
from $-1$ to 1 in SU(2) (e.g. around the third axis). This is clearly, a topologically stable configuration, albeit the stability is of the $Z_2$ type. Correspondingly, the winding ansatz takes the form
\beq
q^{kA}\longrightarrow\sqrt{
\frac{\xi}{2}}
\left(
\begin{array}{cc}
\,e^{i\alpha}  &  0 \\
0 &  1\\
\end{array}
\right)\quad {\rm or} \quad 
q^{kA}\longrightarrow\sqrt{
\frac{\xi}{2}}
\left(
\begin{array}{cc}
1 &  0 \\
0 &  \,e^{i\alpha}\\
\end{array}
\right),
\label{12}
\eeq
depending on whether we use the combination of generators $T_{{\rm U}(1)} + T^3_{{\rm SU}(2)}$
or $T_{{\rm U}(1)} - T^3_{{\rm SU}(2)}$. 

It is clear, that the ansatz (\ref{12}) breaks the color-flavor locked SU(2) down to U(1).
The particular way of embedding is unimportant. Instead of $T^3$ we could have chosen any other generator of SU(2).
In other words, the existence of the string (\ref{12}) implies the existence of the whole family of strings parametrized by
SU(2)/U(1) moduli. The theory of the moduli fields on the string world sheet is the sigma model on the SU(2)/U(1) 
coset space. This is the celebrated CP(1) model. It is asymptotically free in the UV and strongly coupled in the IR. 
Since the bulk theory has eight supercharges and the string is 1/2 BPS saturated, the world-sheet model has four supercharges. In other words, its supersymmetry is ${\mathcal N}=(2,2).$ The existence of the orientational moduli means that the flux through the string does not have a preferred orientation inside SU(2). This is a genuinely non-Abelian string. Non-Abelian strings are formed if all non-Abelian bulk degrees of freedom participate are equally operative at the scale of string formation

\begin{figure}[h]
\epsfxsize=6cm
\centerline{\epsfbox{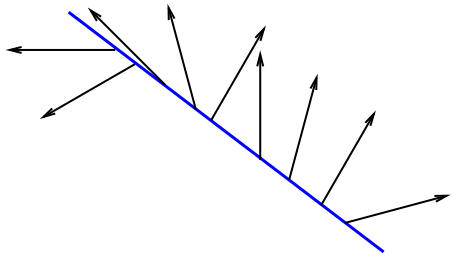}}
\caption{\small
O(3) sigma model on  the string world sheet.}
\label{rota}
\end{figure}

Since the CP(1) model is equivalent to O(3) (see e.g. \cite{SA}), the orientational moduli can be represented as
a unit vector (in the ``isospace") attached to every point of the string and allowed to fluctuate freely see Fig. \ref{rota}.

If the tension of the ANO string is $4\pi\xi$, the tension of the non-Abelian string is $2\pi\xi$, in the U(2) bulk theory. Thus, the ANO string is, in a sense, composite.

For bulk theories with the U$(N)$ gauge group the world sheet theory on the non-Abelian string is given
by CP$(N-1)$ models \cite{AHDT}.

\section{Supersymmetry in the bulk and BPS strings}
\label{sitb}

The degree of supersymmetry of BPS-saturated \cite{BPS} non-Abelian strings is determined by supersymmetry in the bulk.
Thus, ${\mathcal N} =2$ bulk theories support  ${\mathcal N}  =(2,2)$ models on their world sheet \cite{AHDT},
while ${\mathcal N} =0$ bulk theories give rise to non-supersymmetric non-Abelian strings (e.g. \cite{odi}). The most interesting type of strings -- heterotic -- appear in the ${\mathcal N} =1$ bulk theories \cite{Edalati:2007vk,Shifman:2008wv,BSY}. In this case the world-sheet Lagrangian possesses ${\mathcal N}  =(0,2)$ supersymmetry (it can be minimal or nonminimal), which is usually further spontaneously broken due to an appropriate Goldstino field on the world sheet.

Nonsupersymmetric and (2,2) supersymmetric two-dimensional sigma models are thoroughly studied, see e.g. the
review \cite{Pere}. As far as heterotic (0,2) models are concerned,
till recently only some general aspects have been discussed \cite{west,Witten:2005px,bai2,bai3,Jia,bai4}. However, emergence of these
theories on the string world sheet gave a strong impetus for further studies, see e.g. \cite{L1,L2,C1,C2,C3,C4}.

\section{Basic models}
\label{bamo}

In the vast majority of examples studied so far, the world-sheet theories on non-Abelian strings are
various versions of CP$(N-1)$ models: with or without twisted masses, with or without extra fields, with or without supersymmetry, and their extensions such as the so-called $zn$ and weighted CP$(N,M)$ models. All these models have two (sometimes even three) distinct representations. In this section we will briefly discuss these representation using the simplest example:
non-supersymmetric and (2.2) supersymmetric CP$(N-1)$ without twisted mass.

\subsection{Geometric  formulation}
\label{41}

A generic Lagrangian of any  sigma-model with the K\"ahlerian target space
is
\begin{equation}
\label{eq:kinetic}
{\cal L}_{{\rm CP}(N-1)}
=G_{i\bar j} \, \partial^\mu \bar\phi^{\,\bar j}\, \partial_\mu\phi^{i}\,,
\end{equation}
where $G_{i\bar j}$  is the K\"ahler metric,
$$
G_{i\bar j}=\frac{\partial^{2} K(\phi,\,\bar\phi)}{\partial \phi^{i}\partial \bar\phi^{\,\bar j}}
$$
and $K(\phi, \bar\phi)$ is the K\"ahler potential.
For the  CP($N\!-\!1)$ model one can choose the following K\"ahler potential:
\begin{equation}
\label{eq:kahler}
K=\frac{2}{g_{0}^{2}}\log\left(1+\sum_{i,\bar j=1}^{ N-1}\bar\phi^{\,\bar j}\delta_{\bar j i}\phi^{i}\right)\,,
\end{equation}
corresponding  to the so-called round Fubini-Study metric.  The bare coupling constant is denoted by $g_0^2$.

 It is not difficult to supersymmetrize the model (\ref{eq:kinetic}) and (\ref{eq:kahler}). Its ${\mathcal N} = (2.2)$ generalization can be written as  \cite{WessBagger}
\beqn
\label{skinetic}
{\cal L}_{{\mathcal N}=(2,2)}=\!\int\! {d}^{4 }\theta K(\Phi, \bar\Phi)
&=&G_{i\bar j} \left[\rule{0mm}{6mm}\partial^\mu \bar\phi^{\,\bar j}\, \partial_\mu\phi^{i}
+i\bar \psi^{\bar j} \gamma^{\mu} {\mathcal D}_{\mu}\psi^{i}\right]
\nonumber\\[2mm]
&-&\frac{1}{2}\,R_{i\bar jk\bar l}\,(\bar\psi^{\bar j}\psi^{i})(\bar\psi^{\bar l}\psi^{k})\,,
\eeqn
where $\Phi^{i}$ and $\bar\Phi$ are the chiral and antichiral superfields 
\beq
\Phi^{i}(x^{\mu}+i\bar \theta \gamma^{\mu} \theta),\qquad \bar\Phi^{\,\bar j}(x^{\mu}-i\bar \theta \gamma^{\mu} \theta)
\label{12p}
\eeq
of which the lowest components are $\phi^{i}$ and $\bar\phi$
(see e.g. \cite{SA}),
$R_{i\bar jk\bar l}$ is the Riemann tensor, 
\beq {\mathcal D}_{\mu}\psi^{i}=
\partial_{\mu}\psi^{i}+\Gamma^{i}_{kl}\partial_{\mu} \phi^{k}\psi^{l}
\label{12pp}
\eeq
is the covariant derivative, $\Gamma^{i}_{kl}$ are the Christoffel symbols, and we use the notation
$\bar \theta=\theta^{\dagger}\gamma^{0}$, $\bar \psi=\psi^{\dagger}\gamma^{0}$
for the fermion objects.   The $\gamma$ matrices  are chosen as 
\beq
\gamma^{0}=\gamma^t=\sigma_2\,,\qquad \gamma^{1}=\gamma^z = i\sigma_1\,,\qquad \gamma_{5} 
\equiv\gamma^0\gamma^1 = \sigma_3\,.
\label{sieeight}
\eeq
For 
CP$(N-1)$ target space, as for any
symmetric manifold, the Ricci-tensor $R_{i\bar j}$ is proportional to the metric, see Eq. (\ref{16}) below.
Both versions of 
this model -- supersymmetric and non-supersymmetric are asymptotically free \cite{poly}. In the former case the $\beta$ function is one-loop exact,
\beq
\beta_{{\mathcal N}=(2,2)}\equiv \,\frac{\partial g_0^2}{\partial \log M_{\rm uv} }=  -\, \frac{g^4N}{4\pi}\,.
\eeq
Only bosons contribute at first loop. In non-supersymmetric CP$(N-1)$ model \cite{FF,SA}
\beq
\beta_{{\rm CP}(N-1)}=  -\, \frac{g^4N}{4\pi}\left(
1+\frac{g^2}{2\pi}+ ...
\right),
\label{fite}
\eeq
 where ellipses stand for the third and higher loops.
 
 For completeness, concluding this section let us add a few extra useful expressions,
 \beqn
&& G_{i\bar j}=\frac{2}{g^{2}}\Bigg(\frac{\delta_{i\bar j}}{\chi}-\frac{\bar\phi^{\,i}\phi^{\bar j}}{\chi^{2}}\Bigg)\,,\qquad
\qquad
~G^{i\bar j}=\frac{g^{2}}{2}\chi\Big(\delta^{i\bar j}+\phi^{i}\,\bar\phi^{\,\bar j}\Big)\,, \nonumber\\[2mm]
&& \Gamma^{i}_{kl}=-\frac{\delta^{i}_{k}\,\bar\phi^{\,\bar l}+\delta^{i}_{l}\,\bar\phi^{\,\bar k}}{\chi}\,,
\qquad \qquad\quad ~\Gamma^{\bar i}_{\bar k \bar l}=-\frac{\delta^{\bar i}_{\bar k}\,\phi^{  l}+\delta^{\bar i}_{\bar l}\,\phi^{ k}}{\chi}\,,\nonumber \\[2mm]
&& R_{i{\bar j}k{\bar l}}=-\frac{g^{2}}{2}\Big(G_{i\bar j}G_{k\bar l}+G_{k\bar j}G_{i\bar l}\Big)\,,\quad 
 ~R_{i\bar j}= - G^{k \bar j}R_{i{\bar j}k{\bar l}}=\frac{g^{2}N}{2}\, G_{i\bar j}\,,\nonumber\\[3mm]
&& \chi\equiv1+\sum_{m}^{N-1}\bar \phi^{\, \bar m}\phi^{m}\,.
\label{16}
\eeqn

 \subsection{Gauged formulation}
 \label{gafo}
 
An alternative formulation -- the so-called gauged formulation --
was suggested by Witten \cite{gfw,W93}. Being completely equivalent to the geometric formulation
it is more convenient for the large-$N$ solution of the model.

The CP$(N-1)$ target space is the coset SU$(N)/({\rm SU}(N-1)\times {\rm U}(1)$. In the gauged formulation we build 
the Lagrangian ${\mathcal L}_{{\rm CP}(N-1)}$ starting from an $N$-plet of 
complex bosonic fields $n^i$ where $i=1,2, ..., N$.
The fields $n^i$ are scalar (i.e. spin-0), and 
 are subject to the constraint
 \beq
\bar n_i \, n^i =1\,,
\label{consed}
\eeq
The Lagrangian takes the form
\beq
\label{lala}
{\mathcal L}_{{\rm CP}(N-1)} =\frac{2}{g^2} 
\left| {\mathcal D}_\mu n^i\right|^2 -D \left( n_i^\dagger \, n^i -1\right)\,,
\eeq
where the covariant derivative ${\mathcal D}_\mu$ is defined as
\beq
{\mathcal D}_\mu n^i \equiv \left(\partial_\mu -i A_\mu\right) n^i\,.
\label{copr}
\eeq
Here $A_\mu$ is an auxiliary vector field implementing U(1) gauge invariance, while 
$D$ is an auxiliary real scalar field implementing the constraint (\ref{consed}).
Neither $A_\mu$ nor $D $ have kinetic terms in the Lagrangian (\ref{lala}).

Sometimes it is convenient to rescale the $n$ and $D$ fields as follows:
\beqn
\label{lalap}
{\mathcal L}_{{\rm CP}(N-1)} &=&
\left| {\mathcal D}_\mu n^i\right|^2 -D \left( \bar n_i\, n^i -2\beta \right)\,,
\nonumber\\[2mm]
  \beta &\equiv& \frac{1}{g^2}\,,
\label{lalam}
\eeqn
making the kinetic term canonic. The vacuum expectation value of $D$ will then play 
the role of the $n$-field mass squared.

From Sec. \ref{41} we see that the CP$(N-1)$ target space is parametrized by $2N-2$ real degrees of freedom.
There are $2N$ real degrees of freedom in the $n^i$ fields. The constraint (\ref{consed})
reduces this number to $2N-1$, while the U(1) gauge invariance further reduces it to 
 $2N-2$.

The fields $\phi^i$ of the geometric formulation can be related to $n^i$ (on a particular patch) by singling out
 one of the components of $n^i$, say, $n^N$, and defining
\beq
\phi^i = \frac{n^i}{n^N}\,,\quad i = 1,2, ..., N-1\,.
\label{dvod}
\eeq

\vspace{2mm}

The easiest way to extend the above formalism to ${\mathcal N}= (2,2)$ supersymmetry is to start
from ${\mathcal N}=1$ SQED in four dimensions with $N$ flavors of chiral matter superfields (with one and the same charge), plus the Fayet-Iliopoulos term $\tilde\xi$,
\beq
{\mathcal L } =
 \left\{ \frac{1}{4\, e^2}\int\! d^2\theta \, W^2 + {\rm H.c.}\right\} +
\int \! d^4\theta \sum_{i=1}^N\left(\bar{Q}_ie^V Q^i \right)
-
 \tilde{\xi}
  \int\! d^4 \theta \, V \,,
\label{sqed}
\eeq
where 
$$
Q^i = n^i + \sqrt{2}\,\theta\xi^i +\theta^2 F^i\,.
$$

This theory does not exist in four dimensions due to the chiral anomaly. However, we will use it only as a starting point, 
with the intention of reducing it to two dimensions. In two dimensions it becomes well-defined.
The ${\mathcal N}= (2,2)$ CP$(N-1)$ model is obtained in the limit $e^2\to\infty$. In this limit
both the photon and photino kinetic terms can be dropped, and we obtain
(in components)
\beqn
{\mathcal L}_{{\mathcal N}=(2,2)}
&=&
 \left| {\mathcal D}_\mu n^i\right|^2 - 2|\sigma|^2\, |n^i|^2
- D \left(|n^i|^2 -2\beta
\right)
\nonumber\\[2mm]
&+&
\bar\xi_{jR}\,i{\mathcal D}_L\xi^j_R +\bar\xi_{jL}\,i{\mathcal D}_R\xi^j_L
\nonumber\\[2mm]
&+&\left[\sqrt{2} \sigma\bar\xi_{jR}\xi^j_L +\sqrt{2}\bar n_j\left(\lambda_R\xi_L^j +\lambda_L\xi_R^j\right)+{\rm H.c.}\right].
\label{22}
\eeqn
Here $\sigma = (A_x+iA_y)/\sqrt{2}$ is a part of the superfield $V$ (in the Wess-Zumino gauge), along with
$A_{t,z}$, $D$,  and $\lambda_{R,L}$ (for geometrical conventions see Fig.~\ref{mmmon}). All these fields enter in the Lagrangian (\ref{22}) without kinetic terms.
The latter will be generated, however, at one-loop level, dynamically. The spinor fields $\lambda_{R,L}$
implement the constraint $\bar n_j \xi^j =0$. The constraint $|n^i|^2 =2\beta$ is implemented by the auxiliary $D$ field. The covariant derivative is defined in (\ref{copr}).
Finally, the Fayet-Iliopoulos term in (\ref{sqed}) is related to $2\beta$, namely, 
$\tilde{\xi} \to 2\beta$ (note that in two dimensions the Fayet-Iliopoulos term is dimensionless).

\subsection{CP(1): a special case}

The case $N=2$, when we deal with the CP(1) model,  is special. 
Indeed, the CP(1) target space is isomorphic to 
O(3), implying that the CP(1) model can be formulated in terms of a triplet of real fields $S^a$. The O(3) model, in turn, opens the series of the O$(N)$ models. For $N>3$ the O$(N)$ target space is not K\"ahlerian. Thus, supersymmetrization of the O$(N)$ models with $N>3$ results in ${\mathcal N}=(1,1) $ supersymmetry.

To explicitly pass from CP(1) to O(3) one needs
expressions relating the $\vec S$ fields to the $n^i$ fields. Given the fact that in this case
$n^i$'s are spinors of SU(2) while $\vec S$ is the O(3) vector one can write\beq
S^a = \bar n\, \tau^a\, n\,,\quad a=1,2,3,
\eeq
where $\tau^a$ are the Pauli matrices which satisfy
the Fierz transformation formula 
\beq
\vec \tau_{\alpha\beta}\, \vec\tau_{\delta\gamma} = 2
\delta_{\alpha\gamma}\, \delta_{\delta\beta} - \delta_{\alpha\beta}\, \delta_{\delta\gamma}
\,.
\label{ft}
\eeq
Making use of (\ref{ft})  one concludes that
\beq
\vec S^{\,2} = (\bar n \,n)^2 = 1\,.
\eeq
Thus,
\beq
{\mathcal L}_{{\rm O}(3)} = \frac{1}{2g^2} \, \partial_\mu S^a  \partial^\mu S^a \,,\qquad S^a S^a=1\,,\qquad a=1,2,3.
\label{nsymo}
\eeq
Supergeneralization of (\ref{nsymo}) is straightforward \cite{eddieyoung}. One introduces a triplet of real superfields $\sigma^a(x,\,\theta )$,
\begin{equation}
N^a(x,\,\theta ) =S^a +\bar\theta \chi^a + \frac{1}{2}\,\bar\theta\theta\, F^a\,,
\label{spinn259}
\end{equation}
where $\vec S$ and $\vec F$ are bosonic fields while $\vec \chi$
denotes two-component Majorana fields (the $\theta$ coordinate is also Majorana).

Then the supersymmetric  Lagrangian takes the form  
\begin{eqnarray}
{\mathcal L}_{{\rm O}(3)} 
&=&
\frac{1}{g^2}\,
\int d^2\theta   \left( \frac{1}{2}\bar D_\alpha
N^a
D_\alpha N^a \right) 
\nonumber\\[3mm]
&=&
\frac{1}{2g^2}\,
\left\{\partial^\mu S^a\,\partial_\mu S^a+\frac{i}{2}\, \bar\chi^a
\gamma^\mu \stackrel{\leftrightarrow}{\partial}_\mu\chi^a +\vec F{\,^2}
\right\}\,.
\label{spinn260}
\end{eqnarray}
with  the constraint 
\begin{equation}
N^a(x,\,\theta )\, N^a(x,\,\theta ) =1\,,
\label{spinnn261}
\end{equation}
which replaces the nonsupersymmetric version of this constraint $\vec S^{\, 2} =1$.

Decomposing (\ref{spinnn261}) in components we get
\begin{equation}
\vec S^{\,2} =1\,,\qquad \vec S\vec\chi =0\,,\qquad \vec F\vec S =\frac{1}{2} \left(\bar\chi^a\chi^a\right)\,.
\label{spinnn262}
\end{equation}
As usual, the $F$ term enters with no derivatives. Eliminating $F$ by virtue of the equations of motion one 
obtains \cite{eddieyoung,nsvzobz}
\begin{equation}
{\mathcal L}=
\frac{1}{2g^2}\,
\left\{\partial^\mu S^a\,\partial_\mu S^a+\frac{i}{2}\, \bar\chi^a
\gamma^\mu \stackrel{\leftrightarrow}{\partial}_\mu\chi^a +\frac{1}{4}\, 
\left(\bar\chi^a\chi^a\right)^2
\right\}\,,
\label{spinn267}
\end{equation}
plus the first two constraints in Eq.~(\ref{spinnn262}).

The global O$(3)$ symmetry is explicit in this Lagrangian. Moreover, $(1,1)$ supersymmetry is built in. 
In fact, supersymmetry of this model is ${\mathcal N}=(2,2)$ due to the K\"ahlerian nature of the target space 2-sphere \cite{Zuminosigma}.

A special nature of the CP(1) target space manifests itself in the fact that for the CP$(N-1)$ with $N>2$ minimal heterotic $(0,2)$ models do not exist \cite{Moore}, while it does exists for CP(1), see Sec. \ref{geformu}.

\section{Witten's large-\boldmath{$N$} solution}

In this section we will briefly discuss large-$N$ solutions of the simplest two-dimensional models
emerging on the world sheet of non-Abelian strings. Massless non-supersymmetric and \ntwot supersymmetric
CP$(N-1)$ models were solved at large $N$ by Witten \cite{gfw}.
 
 \subsection{Non-supersymmetric CP\boldmath{$(N-1)$}}
\label{nsc}

Let us turn to (\ref{lalam}) rescaling the coupling constant \`a la 't Hooft to make explicit the
$N$ dependence,
\beq
{\mathcal L}_{{\rm CP}(N-1)} =  \left(\partial_\alpha +iA_\alpha\right)\bar{n}_i
\left(\partial^\alpha-iA^\alpha\right)  {n}^i
-D \left(\bar{n}_i  n^i -\frac{N}{\lambda}\right)^2
\,,
\label{cpconsPP}
\eeq
where 
\beq
\lambda \equiv \frac{g^2N}{2}\,.
\label{cpconsSS}
\eeq
First, we study the vacuum structure of this model.
Note that the Lagrangian (\ref{cpconsPP}) is quadratic in the $n^i$ fields; therefore these fields can be integrated out,
\beq
Z = \int {\mathcal D} A_\alpha \, {\mathcal D} D
\exp\left\{  -N{\rm Tr}\ln \left[-\left(\partial_\alpha -iA_\alpha \right)^2 - D\right]
+ i\frac{N}{\lambda} \int d^2 x\,  D \right\}\,.
\label{cpcons2}
\eeq
The Lorentz invariance of the theory tells us that if the saddle point exists it must be achieved
at an $x$ independent value of $D$. Hence  
we can treat $D$ as a constant, vary with respect to $D$, and
require the result to vanish. The same Lorentz invariance tells us that at the saddle point $A_\alpha =0$.
In this way we arrive at the following equation:
\beq
\frac{i}{\lambda} + \int\frac{d^2 k}{4\pi^2}\, \frac{1}{k^2 -  D} = 0\,,\qquad 
\frac{1}{\lambda} -\frac{1}{4\pi}\log \frac{M_{\rm uv}^2}{D}\,.
\label{cpcons3}
\eeq
The integral in (\ref{cpcons3}) is logarithmic and diverges in the ultraviolet, therefore we cut it off
at $M_{\rm uv}^2$. In this way, starting from (\ref{cpcons3}),  we arrive at the equation
\beq
D_{\rm vac} \equiv m^2 =  M_{\rm uv}^2 \, e^{-4\pi/\lambda} =  M_{\rm uv}^2 \, e^{-8\pi/Ng^2}\equiv\Lambda^2 \,.
\label{trse}
\eeq
The assumption of existence of the saddle point is confirmed {\em a posteriori}. 
The $n$-quanta mass $m$ is a physical parameter. Therefore, the right-hand side of (\ref{trse}) is 
renormalization-group invariant,  $\Lambda^2$,  the dynamical scale parameter of the CP$(N-1)$ model. 
This is in full agreement with the first coefficient of the $\beta$ function in (\ref{fite}).
The second coefficient is invisible to the leading order in $1/N$.

Integrating the second equation in (\ref{cpcons3}) over $D$ one readily reconstructs the effective potential as a function of $D$,
\beq
V_{\rm eff} =\frac{N}{4\pi}\, D\log\frac{D}{e\,m^2}\,, \quad 
e=2.718...\,.
\label{veff}
\eeq

From the large-$N$ solution one can see that the constraint $\bar n_i n^i =1$ is lifted and the  massive $n$ particles form  a full $N$-plet. The $n$-mass is given in (\ref{trse}). Moreover, it is not difficult to 
see that the field 
$A_\mu$  acquires kinetic terms and become dynamical.  
Expanding the effective action (\ref{cpcons2}) around the saddle point, one
can easily check that cubic and higher orders in $D$ and $A$ are suppressed by powers of $1/\sqrt N$.
The linear term of expansion vanish. This is the essence of Eq.~(\ref{cpcons3}).
We will  focus on the quadratic terms of expansion.

\begin{figure}[h]
\centerline{\includegraphics[width=3in]{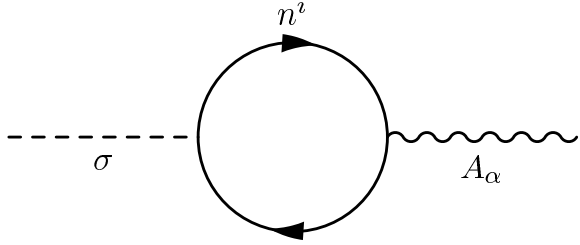}}
\caption{\small The vanishing of the   $D\,A_\alpha$ mixing term  
in the effective Lagrangian.}
\label{cpc2}
\end{figure}

\begin{figure}[h]
\centerline{\includegraphics[width=3in]{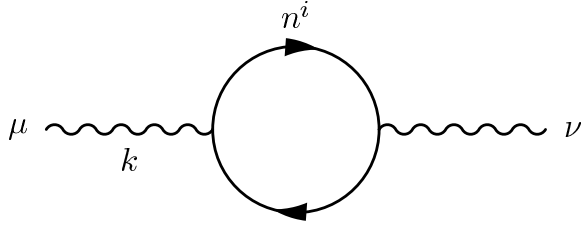}}
\caption{\small $O(A^2)$  terms 
in the effective Lagrangian.}
\label{cp3}
\end{figure}

It is not difficult to check (see e.g. \cite{SA}) that the cross term of the $D A$ type also vanish
(see Fig.~\ref{cpc2}).
Therefore, we need only consider the terms quadratic in $A$, see 
Fig.~\ref{cp3}. 
A straightforward computation yields for the $A^2$ term\,\footnote{The $O(k^4)$, $O(k^4)$, and so terms  can be ignored since they have no impact on the position of the pole at $k^2=0$ of the photon Green's function.  }
\beq
\frac{N}{12\pi\, m^2}\, \left(-g_{\mu\nu} k^2 + k_\mu k_\nu\right)\left(1+O(k^2/m_n^2)\right). 
\label{cpcons7}
\eeq

This expression is automatically transversal, as expected given  the U(1) gauge invariance
of (\ref{cpconsPP}). The $O(k^2)$ term in (\ref{cpcons7})
represents the standard kinetic term $F_{\mu\nu}^2$
of the photon field, more exactly, 
\beq
 -\frac{N}{48\pi\, m^2} F_{\mu\nu}F^{\mu\nu}\,.
 \label{cpcons8}
\eeq
 
It is convenient to rescale the $A$ field to make its kinetic term
 (\ref {cpcons8}) canonically normalized. Upon this rescaling the effective Lagrangian takes the form
\beq
{\mathcal L}_{\rm eff}  =
-\frac{1}{4} F_{\mu\nu}^2+
 \left(\partial_\alpha +i e_n A_\alpha\right)\bar{n}_i
\left(\partial^\alpha -i e_n A^\alpha\right)  {n}^i
- m^2\, \bar{n}_i  n^i 
\,,
\label{cpcons9}
\eeq
where the electric charge of the $n$ quanta $e_n$ is
\beq
e_n \equiv m\,\sqrt{\frac{12\pi}{N}}\,.
\label{cpcons10}
\eeq
It has dimension of mass, which is the correct dimension of the electric charge in two-dimensional theories.
Moreover, one should stress that at large $N$ the electric charge becomes small,
$e_n/m \ll 1$, which implies, in turn, weak coupling.

Emergence of the massless gauge U(1) field ensures the presence of the Coulomb
potential  between charges states. In two dimensions static Coulomb potential is a linear rising potential.
It leads to the confinement of kinks which carry electric charge \cite{gfw}. Therefore this phase of the 
theory is called  Coulomb/confining phase.

 \subsection{Supersymmetric CP\boldmath{$(N-1)$}}
\label{sc}

It is easy to generalize the large-$N$ solution of Sec. \ref{nsc} to ${\mathcal N}=(2,2)$ model \cite{gfw,W93}.
The Lagrangian (\ref{22}) is quadratic in both, the $n$ fields and their fermion superpartners $\xi$. Therefore, they can be integrated out exactly. 
Note that the auxiliary fields $A$, $\sigma$ and $\lambda_{L,R}$ form a supermultiplet. Hence, it is sufficient to find
the kinetic term and mass for one of them in order to determine them all, provided that supersymmetry is unbroken.
As we will see momentarily, it is indeed unbroken. 

As in (\ref{cpcons3}) we set $A_\mu =0$, and then integrate out $n^i$ and $\xi^i$. This yields
\beq
\frac{{\rm Det}\left(\rule{0mm}{4mm} -\partial_\alpha^2 - 2 |\sigma|^2
\right)^N}{{\rm Det}\left(\rule{0mm}{4mm}-\partial_\alpha^2 - D- 2 |\sigma|^2
\right)^N}\,\,.
\label{detdet}
\eeq
 The denominator comes from the boson loop while the numerator from the fermion loop. It is obvious
 that supersymmetric vacuum (with $E_{\rm vac} =0$) is attained at $D=0$, when the ratio of 
 the determinants in (\ref{detdet}) reduces to unity. 
 
 The above conclusion is confirmed by an explicit calculation of  the effective potential, the analog of (\ref{veff}),
 \beq
V_{\rm eff} =\frac{N}{4\pi}\left[\left(D+2|\sigma|^2
\right)\log \frac{D+2|\sigma|^2}{m^2} \, - D - 2|\sigma|^2\log\frac{2|\sigma|^2}{m^2}
\right]\,,
\label{veffp}
\eeq
where we carried out renormalization using the analog of  (\ref{trse}),
\beq
2|\sigma_{\rm vac}|^2 \equiv m^2 =  M_{\rm uv}^2 \, e^{-8\pi/Ng^2}\,.
\label{trsep}
\eeq
 The vacuum values 
of $D$ and $|\sigma|$ are obtained through minimization, i.e. by differentiating $V_{\rm eff}$ in (\ref{veffp})
over $D$ and $2|\sigma|^2$,
\beqn
&& \log \frac{D+2|\sigma|^2}{m^2}=0\,,
\nonumber\\[2mm]
&& \log \frac{D+2|\sigma|^2}{m^2}-\log\frac{2|\sigma|^2}{m^2} =0\,.
\label{vas}
\eeqn
As was mentioned, the mass of the $\xi$ field is the same as as that of $n$, due to supersymmetry.

The kinetic term of the $A_\mu$ field and its superpartners is dynamically generated in much the same way as
in Sec. \ref{nsc}. A crucial difference is that now the photon field $A_\mu$ acquires a nonvanishing 
(albeit small) mass through
the Schwinger mechanism: the massless fermion loop shifts the pole in the photon propagator away from zero. This was noted already in 1979 \cite{gfw}.
Needless to say, all superpartners of the photon field receive the same mass.

Consequences of massless vs. massive photon in two dimensions are radically different. Massless photons in two-dimensions (non-supersymmetric CP$(N-1)$) lead to confinement of charged particles, while massive photons (supersymmetric CP$(N-1)$) do not confine. In one-to-one correspondence with this 
is the existence of $N$ degenerate vacua in the non-confining case.
In the confining case (i.e. massless photon) one of these vacua remains genuine while the remaining $N-1$ are uplifted and become quasistable states.

\section{Twisted masses}
\label{twima}

The so-called twisted masses is the only mass deformation of the ${\mathcal N} =(2,2)$ model which preserves 
supersymmetry. The essence of this deformation is as follows \cite{AGF}. One starts from four-dimensional
CP$(N-1)$ model and couples $N-1$ conserved U(1) currents of this model to background gauge four-potential $A_\mu$. Then one reduces the model to two dimensions $t$ and $z$ simultaneously declaring
the background fields $A_x$ and $A_y$ (Fig. \ref{mmmon}) to be nonvanishing constants. The twisted masses
$\mu$ and $\bar\mu$ are proportional to $A_x\pm i A_y$.

In the geometric formulation of Sec. \ref{41} the formal procedure can be described as follows. The theory (\ref{eq:kinetic}) can be interpreted as an ${\cal N}=1$ theory of $N-1$ chiral superfields 
in four dimensions.  The theory possesses  $N-1$  U(1) isometries 
parametrized by $t^{a}$, $a=1,\ldots, N-1$.
The Killing vectors of the isometries can be expressed via derivatives of the Killing 
potentials $D^{a}(\phi, \phi^{\dagger})$,
\begin{equation}
\label{eq:KillD}
\frac{{d}\phi^{i}}{{  d}\,t_{a}}=-iG^{i\bar j}\,\frac{\partial D^{a}}{\partial \bar\phi^{ \,\bar j}}
\,,\qquad 
\frac{{d}\bar\phi^{\,\bar j}}{{  d}\,t_{a}}=iG^{i\bar j}\,\frac{\partial D^{a}}{\partial \phi^{i}}\,.
\end{equation}
This defines the U(1) Killing potentials up to additive constants.

The isometries are
 evident from the expression (\ref{eq:kahler}) for the K\"ahler potential, 
\begin{equation}
\label{eq:iso}
\delta\phi^{i}=-i\delta t_{a} (T^{a})^{i}_{k}(\phi)^{k}\,,\qquad 
\delta\bar\phi^{\,\bar j}=i\delta t_{a}(T^{a})^{\bar j}_{\bar l}\bar\phi^{\,\bar l}\,,
\qquad a=1,\ldots, N-1\,,
\end{equation}
(together with the similar variation of fermionic fields),
where the  generators $T^{a}$ have a simple diagonal form,
\begin{equation}
(T^{a})^{i}_{k}=\delta^{i}_{a}\delta^{a}_{k}\,, \qquad a=1,\ldots,N-1\,.
\end{equation}
 The explicit form of the Killing potentials $D^{a}$ in CP$(N\!-\!1)$ with the Fubini--Study metric is
\beq
\label{eq:KillF}
D^{a}=\frac{2}{g_{0}^{2}}\,\frac{\bar\phi\,T^{a}\phi}{1+\bar\phi\,\phi}\,,
\qquad a=1,\ldots,N-1\,.
\eeq
Here we use the matrix notation implying that $\phi$ is a column $\phi^{i}$ and 
$\bar\phi$ is a row $\bar \phi^{\, \bar j}$.

The isometries allow us  to introduce an interaction with $N-1$ {\em distinct} background
U(1) gauge 
superfields $V_{a}$ by modifying the K\"ahler potential (\ref{eq:kahler}) in a gauge invariant way,  
\begin{equation}
K(\Phi, \bar\Phi)\to
\tilde K(\Phi, \bar\Phi,V) 
=
\label{eq:mkahlerp}
\frac{2}{g_{0}^{2}}\log\big(1+\bar\Phi\,{\rm e}^{V_{a}T^{a}}\Phi\big)\,.
\end{equation}
where
\beq
V_{a}=-\mu_{a}\bar \theta(1+\gamma_{5})\theta -\bar \mu_{a}\bar \theta(1-\gamma_{5})\theta\,.
\end{equation}
Thus, in our notation the complex masses  $m_{a}$ are linear combinations of the
constant U(1) gauge potentials,
\beq
m_{a}=A^{a}_{y}+iA^{a}_{x}\,,\qquad \bar m_{a}=m_{a}^{*}=A^{a}_{y}-iA^{a}_{x}\,.
\end{equation}

Passing to two dimensions we assume, of course,  that there is no dependence
on $x$ and $y$ in the chiral fields.  It gives us the Lagrangian with the twisted masses 
included \cite{AGF},
\beqn
{\mathcal L}_{m}= \int {d}^{4 }\theta \,K_{m}(\Phi, \Phi^{\dagger},V)
&=& G_{i\bar j}\, g_{MN}\left[{\cal D}^M\! \phi^{\dagger\,\bar j}\, {\cal D}^{N}\!\phi^{i}
+i\bar \psi^{\bar j} \gamma^{M}D^{N} \!\psi^{i}\right]\nonumber\\[3mm]
&-&\frac{1}{2}\,R_{i\bar jk\bar l}\,(\bar\psi^{\bar j}\psi^{i})(\bar\psi^{\bar l}\psi^{k})\,,
\label{eq:mtwist}
\eeqn
where summation over $M$ includes, besides $M=\alpha=0,1$, also 
$M=+,-$. 
The  metric $g_{MN}$ and extra $\gamma$ matrices are
\begin{equation}
\label{eq:metric}
g_{MN}=\left(\begin{array}{crrr}1& 0& 0 & 0 \\0 & -1 & 0 & 0 \\[1mm]0 & 0 & 0 & -\frac 1 2 \\[1mm]0 & 0 & -\frac 1 2 & 0\end{array}\right),\qquad
\gamma^{+}=-i(1+\gamma_{5})\,,\quad
\gamma^{-}=i(1-\gamma_{5})\,.
\end{equation}
The gamma-matrices satisfy the following algebra:
\beq
\bar\Gamma^{M}\Gamma^{N}+\bar\Gamma^{N}\Gamma^{M}=2 g^{MN}\,,
\eeq
where the set $\bar\Gamma^{M}$ differs from $\Gamma^{M}$  by interchanging of
the $+,-$ components, $\bar\Gamma^{\pm}=\Gamma^{\mp}$.
The gauge covariant derivatives ${\cal D}^M$ are defined as
\beqn
&& {\cal D}^{\alpha}\phi
=
\partial^{\alpha}\phi\,,\qquad {\cal D}^{\alpha}\bar\phi =\partial^{\alpha}\bar\phi\,,
\nonumber\\[2mm]
&&{\mathcal D}^{+}\phi
=
 -\bar \mu_{a}T^{a}\phi\,,
\quad  {\mathcal D}^{-}\phi=\mu_{a}T^{a}\phi\,,
\nonumber\\[2mm]
&& {\mathcal D}^{+}\bar\phi =\bar\phi T^{a}\bar \mu_{a}\,,
\quad  {\mathcal D}^{-}\bar\phi =-\bar \phi T^{a} \mu_{a}\,,
\eeqn
and similarly for ${\cal D}^{M}\psi$, while the general covariant derivatives $D^{M}\psi$'s are
\beq
D^{M}\psi^{i}=
{\mathcal D}^{M}\psi^{i}+\Gamma^{i}_{kl}\,{\cal D}^{M}\! \phi^{k}\,\psi^{l}\,.
\eeq

In the geometrical formulation we have $N-1$ complex twisted mass parameters. Introduction of the
twisted masses in the gauged formulation will be discussed in Part II, Secs. \ref{tmz} and \ref{tmzp}. 
In the gauged formulation there are $N$ complex twisted mass parameters $m_i$ related to $\mu^a$,
\beq
\mu^i = m_i - m_N\,,\qquad i=1,2, ..., N-1\,.
\eeq
 (see (\ref{dvod}))
 and subject to the constraint
 \beq
 \sum_{i=1}^N m_i = 0\,.
 \eeq
 One of our tasks in what follows is the study of the phase diagram of the two-dimensional model on the
 string world sheet. To this end it is convenient to have a discrete symmetry. A $Z_N$ symmetry is guaranteed if  the mass parameters are adjusted as
 \beq
 m_j = m_0\,\exp\left(\frac{2\pi i\,j}{N}\right),\qquad j = 1,2, ... , N\,.
 \label{61}
 \eeq
 Such a choice is referred to as $Z_N$ symmetric. It is always assumed in what follows if not stated to the contrary.
 
 Note that $m_0$ can be chosen to be real and positive. Then $m_N$ is real and positive too. Alternatively, if $m_0 =|m_0|\exp\left(-2\pi i/N
 \right)$, then $m_1$ is real and positive. 
 
 \newpage
 
 \newpage

\rule{0mm}{9mm}

\vspace{1cm} 

\begin{center}
{\Huge Part II\\[6mm]

Travel Guide}
\end{center}

\vspace{2.cm}

\begin{center}
\includegraphics[width=3.5in]{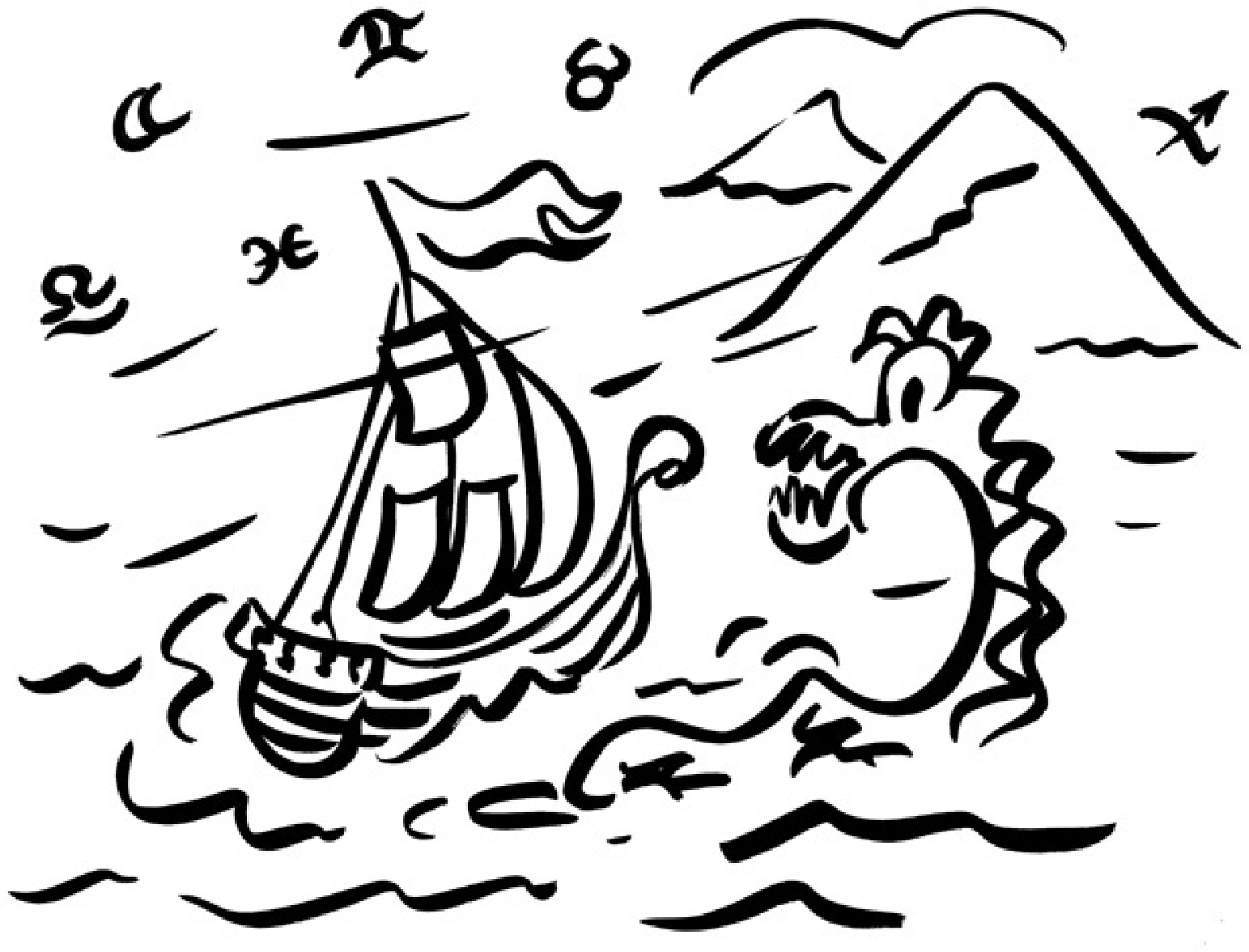}
\end{center}

\vspace{1.5cm}

\noindent
In Part II we will use Euclidean conventions. The Euclidean action reduces to the energy functional for static fields.

\newpage

\section{Large-$N$ solutions with twisted masses}
\label{lnswtm}

In this section we will briefly review those two-dimensional sigma models that are in the limelight ever since the
discovery of the non-Abelian strings.

\subsection{CP\boldmath{$(N-1)$} with \boldmath{$Z_N$} symmetric masses} 
\label{tmz}

As a world-sheet model in the non-supersymmetric context, CP{$(N-1)$} with twisted masses was discussed
in \cite{odi}, and its large-$N$ solution in the $Z_N$ symmetric case was found in \cite{sgmsay}.

In the gauged formulation the Lagrangian has the form
\beq
{\mathcal L}  =  
|{\mathcal D}_{\alpha} n^{i}|^2
+D \left( | n^{i}|^2-2\beta\right)
+ \sum_{i =1}^N\left|(\sigma - m_i)n^i\right|^2
 \,,
\label{model}
\eeq
were ${\mathcal D}_{\alpha}=\partial_{\alpha}-
iA_{\alpha}$, the mass parameters $m_i$ are defined in Eq. (\ref{61}), and $2\beta$
is the bare coupling constant, see Eq. (\ref{lalam}).

Intuitively it is clear that the structure of the solution depends on the ratio of $m$ and
the dynamical scale $\Lambda$ generated in this theory. As we will see below, there are two distinct cases --
the Higgs and the Coulomb/confining  phases -- in this theory at large and small $|m_0/\Lambda |$, respectively. 

In the Higgs phase the field $n^{i_0}$ develops a VEV.
One can always choose $i_0=1$ and denote $n^{i_0}=n^1\equiv n$. 
There are $N$ equivalent choices, $N$ vacua. This corresponds to the spontaneous breaking of $Z_N$.
Setting the background $A_\alpha$ field to zero, as in Sec. \ref{nsc},  and integrating out all $n_i$ except $n^{i_0}
=n$
we arrive at
\beqn
 {\mathcal L}_{\rm eff} &=&  
|\partial_{\alpha} n|^2
+\left(D +  |\sigma - m_1|^2\right)|n|^2
\nonumber\\[3mm]
&+&\frac1{4\pi}\, \sum_{i=2}^{N}\left(D
+|\sigma-m_{i}|^2\right)
\left[1-\log\, {\frac{D +|\sigma-m_{i}|^2}{\Lambda^2}}\right]
\nonumber\\[3mm]
&+& 
\frac1{4\pi}\,c\,
\sum_{i=2}^{N}|\sigma -m_{i}|^2 
\label{effaction}
\eeqn
where
\beq
c=\frac{1}{N}\,
\sum_{i=2}^{N}\left(1-\frac{m_{i}}{m_1}\right)
\log\, {\frac{ |m_{i}-m_1|^2}{\Lambda^2}}\,,
\label{c}
\eeq
and we used the renormalization condition 
\beq
\frac{2}{g_0^2}=\frac{N}{4\pi}\, \ln\, {\frac{M_{\rm uv}^2}{\Lambda^2}}\,.
\label{65}
\eeq
This condition introduces the dynamical scale $\Lambda$ through dimensional transmutation, just like 
in Sec. \ref{nsc}.

Minimizing this effective potential with respect  to  $D$, $n$ and
$\sigma$ we determine the vacuum values of these parameters. It is not surprising that $n_{\rm vac}$ 
turns out to be exactly as it follows from the renormalized  constraint $| n^{i}|^2=2\beta$ in the Higgs phase,
\beq
n_{\rm vac}= \left(\frac{N}{2\pi}\, \log\, \left| {\frac{m_0}{\Lambda}}\right|
\right)^{1/2}
\label{nva}
\eeq 
while
\beq
D _{\rm vac}=0\,,\qquad \sigma _{\rm vac} = m_1\,.
\label{nvb}
\eeq
It is also obvious that there are $N$ vacua corresponding to cyclic permutation. In each of them the $Z_N$ symmetry is spontaneously broken.  

\begin{figure}
\epsfxsize=8cm
\centerline{\epsfbox{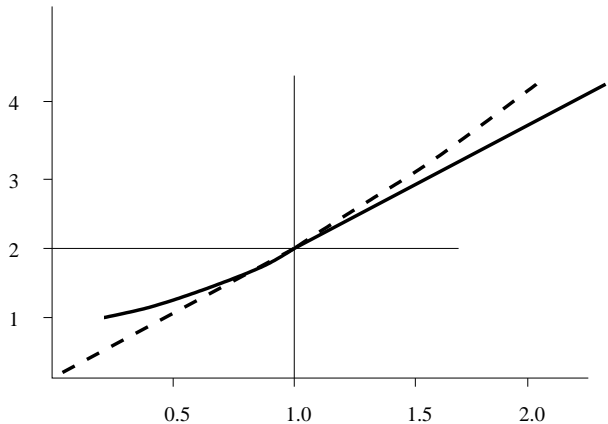}}
\caption{\small 
Normalized vacuum energies $(4\pi E_{\rm vac}/N\,\Lambda^2)$ versus 
$m^2_0/\Lambda^2$. The solid line shows the actual vacuum energy,
while dashed lines correspond to a formal extrapolation of the Higgs and 
Coulomb/confinement
vacuum energies to unphysical values of $m$ below and above 
the phase transition point,
respectively.}
\label{fig:energies}
\end{figure}

Substituting (\ref{nva}) and (\ref{nvb}) in (\ref{effaction}) we obtain the vacuum energy density,
\beq
E_{\rm Higgs\,\, vac} =\frac{N}{2\pi}\, m^2_0\, ,
\label{higgsenergy}
\eeq
where the parameter $m_0$  is assumed to be real and positive (see the bold line in Fig. \ref{fig:energies}).
This formula is valid at
\beq
m_0\ge\Lambda\,.
\label{higgsregion}
\eeq
The Higgs phase has a clear-cut meaning at large $m_0$. The
above result is compatible with intuition.
We will see momentarily that the lower bound of the allowed domain,
$m_0=\Lambda$, is the phase transition point (presumably, the phase transition is of the second order).

Now let us discuss the Coulomb/confining phase. At small $|m_0|$
\beq
\sigma_{\rm vac}=0 \, ,\qquad (n^i)_{\rm vac}=0 \,\,\, {\rm for}\,\,\, {\rm all}\,\, i, \,\,\,  i=1,2, ..., N\,,
\label{cvacp}
\eeq
and
\beq
D_{\rm vac}= \Lambda^2 -m^2_0\, .
\label{cvac}
\eeq
The 
$Z_N$ symmetry remains unbroken. Hence, we deal with a unique vacuum. 
Inspecting  Eq.~(\ref{model})
we conclude that in the saddle point the mass of all $n^i$ quanta
is $\Lambda$, independent of the value of the mass deformation
parameter
$m_0$. 

The vacuum energy in  the Coulomb phase
is obtained by substituting the vacuum values (\ref{cvacp}) and (\ref{cvac})  in 
(\ref{effaction})
and using expression  (\ref{c}) for the value of the constant $c$. In this way one arrives
at 
\beq
E_{\rm Coulomb\,\, vac}
=\frac{N}{4\pi}\,\left \{\Lambda^2 +m^2_0+
m^2_0\log{\frac{m^2_0}{\Lambda^2}}\right \}\,,
\label{coulombenergy}
\eeq
(see the solid line in Fig. \ref{fig:energies}).

 At the point of the phase transition
at $m_0=\Lambda$ the energy densities in the both phases coincide. Moreover,
their first derivatives
with respect to $m^2_0$ at this point coincide
too. The dashed line
corresponds to a formal extrapolation of the Higgs and 
Coulomb/confinement
vacuum energies to ``forbidden" values of $m_0^2$ below and above 
the phase transition point.

To reiterate, at  $m_0\geq \Lambda $, at weak coupling, we have $N$ strictly
degenerate vacua;
the $Z_N$ symmetry is broken. At $m_0\leq \Lambda $ the $Z_N$ symmetry is unbroken, and the vacuum is unique.
 The order parameter which marks
these vacua is the  VEV  of $n^i$.

Introduction of an additional axion field in this model is discussed in \cite{axion}.

\subsection{Supersymmetric CP\boldmath{$(N-1)$} with \boldmath{${\mathcal N}=(2,2)$} } 
\label{tmzp}

Two-dimensional  CP$(N-1)$ models with twisted masses and with ${\mathcal N}=(2,2)$ supersymmetry \cite{HaHo}
emerge as effective low-energy theories on
the world sheet of non-Abelian strings in a class of 
four-dimensional ${\mathcal N}=2$ gauge theories with unequal (s)quark 
masses~\cite{AHDT}, for a complete derivation see \cite{BSY}.  In the gauged formulation
the CP$(N-1)$ Lagrangian with the twisted masses (replacing the zero mass limit (\ref{22}))
is (see e.g. \cite{BSY,L2})
 \beqn
{\mathcal L} 
&=&
  \left[
	\left| {\mathcal D} n \right|^2  + 2 \left| \sigma - \frac{m^l}
	{\sqrt{2}} \right|^2 \left| n^l \right|^2
	+ iD \left( \left|n^l \right|^2 - 2\beta \right)\right.
	\nonumber\\[2mm]
&+&
	 \bar\xi_R\, i{\mathcal D}_L \xi_R  +  \bar\xi_L\, i{\mathcal D}_R \xi_L+
	i\sqrt{2}\, \left( \sigma - \frac{m^l}{\sqrt{2}} \right) \,{\bar\xi}_{Rl} \xi_L^l
	+ i\sqrt{2}\, \left( \bar{\sigma} - \frac{\bar{m}^l}{\sqrt{2}} \right) \bar{\xi}_{Ll} \xi_R^l
	\nonumber\\[4mm]
&+&\left.
	\left( i\sqrt{2}\, {\bar\xi_{R} \lambda}_{L}\, \,n
	-  i\sqrt{2}\, \bar{n}\,  \lambda_{R}\, \xi_{L} + {\rm H.c.}\right)
	\rule{0mm}{6mm}\right].
\eeqn

To solve the model in the large-$N$ limit one
 can basically repeat the derivation of Sec. \ref{sc} since the fields $n$ and $\xi$ enter in the Lagrangian bilinearly. Integrating them out yields
\beq
 \frac{
\prod_{i=2}^{N} {\rm det}\, \left(-\partial_{k}^2 
   + \bigl| \sqrt{2}\sigma - m_i \bigr|^2\right)}{
\prod_{i=2}^{N}{\rm det}\, \left(-\partial_{k}^2 +iD
   + \bigl| \sqrt{2}\sigma - m_i \bigr|^2\right)},
\rule{0mm}{10mm}
\label{det}
\eeq
and
we obtain \cite{L2} an analog of the nonsupersymmetric formula  (\ref{effaction})
for the vacuum structure. Above  we integrated over $N-1$
fields $n^{i}$ and $\xi^i$ with $i > 1 $.
The resulting effective action is to be considered as
a functional of $n^1\equiv n$, $D$ and
$\sigma$. 
We will again assume that the twisted masses are
$Z_N$ symmetric, see (\ref{61}). 

The ensuing 
effective Lagrangian is
\beqn
{\mathcal  L} 
&=&
\sum_{i=2}^{N}\frac{1}{4\pi}\left\{\left(iD+\left|\sqrt{2}\sigma-m_i\right|^2\right)
\left( \log\, {\frac{M_{\rm uv}^2}{iD + \bigl|\sqrt{2}\sigma-m_i\bigr|^2}} + 1 \right)
\right.
\nonumber\\[4mm]
&-&
\left. 
\left|\sqrt{2}\sigma-m_i\right|^2
\left(  \log\, {\frac{M_{\rm uv}^2}{\left|\sqrt{2}\sigma-m_i\right|^2}} + 1\right)
\right\},
\label{detr}
\eeqn
Using (\ref{65}) for the bare coupling constant
we can eliminate $M_{\rm uv}$ in a usual way. Then
 the effective potential as 
a function of $n$, $D$ and $\sigma$ fields  takes
the form
\beqn 
	V_{\rm eff}
	&=&
		\left( iD + \left|\sqrt{2}\sigma - m_1\right|^2 \right) |n|^2 
	\nonumber\\[4mm]
	&-& 
	\frac{1}{4\pi}\, \sum_{i=2}^{N} \left( iD + \left |\sqrt{2}\sigma - {m_i}\right|^2 \right)\,
		\log\, \frac{ iD \,+\, \left| \sqrt{2}\sigma - m_i \right|^2} {\Lambda^2}
\nonumber\\[4mm]
	&+& 
	\frac{1}{4\pi}\, \sum_{i=2}^{N} \left| \sqrt{2}\sigma - m_i\right|^2\,
			\log\, \frac{ \left| \sqrt{2}\sigma -  m_i \right|^2 }{ \Lambda^2 }
	+
	\frac{1}{4\pi}\, iD\, (N-1)  \,  .
	\nonumber\\
	\label{Veff22}
\eeqn
Minimization of (\ref{Veff22}) gives two solutions: either
\beq
\label{higgsph22}
	 iD + \left| \sqrt{2}\sigma - m_1 \right|^2 = 0  
\eeq
or
\beq
\label{strongph22}
	 n = 0 \,. 
\eeq
	These two distinct solutions correspond to the weak and strong-coupling regimes of the theory, respectively. 
They are analogous to two phases we observed in Sec. \ref{tmz}. In the case at hand supersymmetry is unbroken but
in both weak and strong-coupling regimes the $Z_N$ symmetry is spontaneously broken, and there are $N$ distinct vacua.
At strong coupling, in the regime $n_{\rm vac}=0$, the order parameter is $\bar\xi_R\xi_L$ and its Hermitean conjugate.

 As usual, supersymmetry suppresses
phase transitions. The passage from weak (large $|m_0|$) and strong (small $|m_0|$) regimes presents a crossover rather than a phase transition. Supersymmetry is preserved in both regimes.
 
 At large $|m_0|$
 
 \beq
D= 0,\qquad \sqrt{2}\sigma_{\rm vac} = m_1,\qquad |n_{\rm vac}|^2= \frac{N}{2\pi}\,\log\,{\frac{m}{\Lambda}} \,.
\label{higgsvac}
\eeq
 This is similar to the Higgs phase in Sec. \ref{tmz}.
 
 \vspace{2mm}
 
 For small $|m_0|$ we have
 \beq
 D=0\,,\qquad n_{\rm vac}=0\,,
 \eeq
 while the vacuum equation on $\sigma$ can be written as
 \beq
\prod_{i=1}^{N}\left|\sqrt{2}\sigma-m_i\right| \,=\,\Lambda^N \,.
\label{Witcond}
\eeq
 
 For the $Z_N$-symmetric masses
Eq. (\ref{Witcond}) can be solved. Say, for even $N$ one can rewrite this equation in the form
\beq
\left|\left(\sqrt{2}\sigma\right)^N - m^N\right| = \Lambda^N \,,
\label{tumvn}
\eeq
due to the fact that with the masses given in (\ref{61})
\beqn
\sum m_i &=& 0\,,
\nonumber\\[2mm]
\sum_{i,j;\,i\neq j}\,\, m_i m_j &=& 0\,,
\nonumber\\[2mm]
&...&
\nonumber\\[2mm]
\sum_{i_1,i_2,...,i_{N-1}} m_{i_1} m_{i_2} ... m_{i_{N-1}} &=& 0\,,\qquad \left(i_1\neq i_2\neq ...\neq i_{N-1}\right).
\eeqn
Equation (\ref{tumvn})  has $N$ solutions (i.e. $N$ distinct vacua),
\beq
\sqrt{2}\sigma ~=~ \left(\Lambda^N+m^N_0\right)^{1/N}\,
\exp\left( \frac{2\pi\,i\, k}{N}
\right), \quad k=1, ..., N,
\label{22sigma}
\eeq

\begin{figure}
\centerline{\input{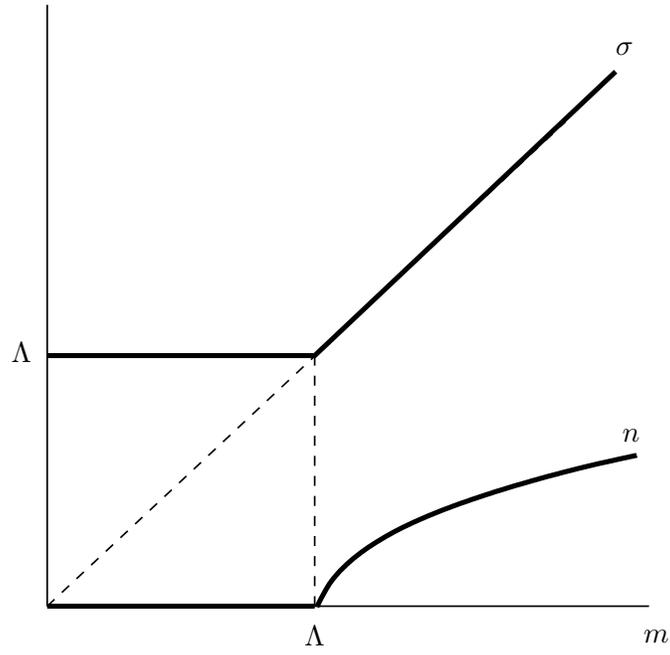}}
\caption{\small Plots of $n$ and $\sigma$ VEVs (thick lines) vs. $m_0$
in the ${\mathcal N} = (2,2)$ CP$(N-1)$  model with the $Z_N$-symmetric twisted masses. }
\label{fig22nsigma}
\end{figure}

The crossover occurs at $m_0=\Lambda$. The width of the crossover domain is not seen in the
leading order in $1/N$. In fact, it is exponentially small in $N$. The transition from weak to strong coupling is depicted in Fig. \ref{fig22nsigma}.

Summarizing, in both regimes 
-- weak and strong coupling -- 
supersymmetry is unbroken and there is no confinement of charged particles
due to the fact that the photon (which becomes dynamical) acquires a mass. The spontaneous breaking of $Z_N$
implies $N$ degenerate vacua.

 \subsection{Curves of marginal stability in \boldmath{$(2.2)$} CP\boldmath{$(N-1)$}  
with $Z_N$ twisted masses}
\label{cms}

An  exact twisted superpotential of the  Veneziano-Yankielowicz  type \cite{VYan} is known to describe  the 
${\mathcal N}=(2,2)$ supersymmetric CP$(N-1)$  model \cite{AdDVecSal,ChVa,W93,HaHo,Dorey}.
Integrating out the fields $n^P$ and $\rho^K$  we obtain
 the following
exact twisted superpotential:
\beq
 {\cal W}_{{\rm CP}(N-1)}(\sigma)= 
\frac1{4\pi}\left\{\sum_{l=1}^N\,
\left(\sqrt{2}\sigma-{m}_l\right)
\,\ln{\frac{\sqrt{2}\sigma-{m}_l}{\Lambda}}
- N \,\sqrt{2}\sigma \right\}\, ,
\label{CPsup}
\eeq
where we use one and the same notation $\sigma$ for the  twisted superfield \cite{W93} and its lowest scalar
component. 
Minimizing this superpotential with 
respect to $\sigma$ we get the equation for the $\sigma$-field VEVs, 
\beq
\prod_{l=1}^N(\sqrt{2}\sigma-{m}_l)
=\Lambda^{N}\,.
\label{sigmaeq}
\eeq
This equation has $N$ roots $\sigma_p$ ($p=1, ..., N)$ associated with $N$ vacua of the
CP$(N-1)$ model.
Note that this {\em exact} equation is a holomorphic version of Eq.~(\ref{Witcond}) which appears in 
the large-$N$ solution. It takes into account
that chiral U(1)$_R$ symmetry is broken by chiral anomaly down to discrete $Z_{2N}$ symmetry.
This is the reason for the  presence of $N$ distinct vacua.

The  masses of the BPS kinks interpolating between the
vacua $\sigma_{p}$ and $\sigma_{p'}$ are given  by the appropriate 
differences of the superpotential (\ref{CPsup}) calculated at distinct roots \cite{HaHo,Dorey,DoHoTo},
\beq
M^{\rm BPS}_{pp'} =
2\left| \rule{0mm}{5mm} {\cal W}_{{\rm CP}(N-1)}(\sigma_{p'})-{\cal W}_{{\rm CP}(N-1)}(\sigma_{p})\,\right|\,,
\qquad p,p'=1,..., \,N\,.
\label{BPSmass}
\eeq
 In addition to kinks, the BPS spectrum of the model
contains elementary excitations with masses given by $|m_l-m_p|$ ($l=1,...,N$ and  $p=1,...,N$).

Due to the presence of branches  in the logarithmic functions in (\ref{CPsup}) each kink comes
 together with a 
tower of dyonic kinks carrying global U(1) charges. The dyonic kinks are reflected in 
(\ref{BPSmass}) through terms
\beq
{\rm integer}\times i\,m_l
\label{dop3}
\eeq
with different $l$ which appear from  the logarithm branches.
We stress that all  these kinks with the imaginary part (\ref{dop3}) in the mass formula (\ref{BPSmass})
interpolate between the same pair of vacua: $p$ and $p'$.

Generically there are  way too many choices  in (\ref{BPSmass}).
Not all of them are realized. Moreover, the kinks present in the quasiclassical domain 
(i.e. at large $\left| m_0\right| $)
decay on the curves of marginal stability (CMS) or form new bound states. Therefore, the quasiclassical
spectrum outside CMS and the quantum spectrum inside CMS (i.e. at small $\left| m_0\right| $) are different.
This phenomenon is referred to as ``wall crossing."
 There exists
a general procedure \cite{KontsSoib} which allows one to determine the full BPS spectrum
starting from the strong coupling spectrum inside CMS. However, this procedure is rather cumbersome.
In certain cases one can use a simpler approach based on analysis 
of various limits. Below we will briefly review the BPS spectra and CMS in 
CP(1) and CP(2) with the $Z_N$ twisted masses \cite{Dorey,SVZw,BSYcms}. 

These CMS
were obtained by matching the weak coupling BPS spectrum found using semiclassical considerations
with the strong coupling spectrum found from mirror representation of the CP$(N-1)$  model \cite{HoVa}. 
The later spectrum  includes only $N$ kinks which become massless at strong coupling.

The strong coupling spectrum of the CP(1) model includes two BPS states with the following masses:
\beq
M_0=|m_D^{\rm CP(1)}+im_1|, \qquad M_1=|m_D^{\rm CP(1)}+im_2|,
\label{strongCP1}
\eeq
where
\beq
\label{mcp1}
	m_D^{\rm CP(1)} =
	\frac{1}{\pi} 
	\left[   2\, \sqrt{ m_0^2 + \Lambda^2\, }
			- m_0 
			    \log\, \frac { \sqrt{ m_0^2 + \Lambda^2\, } + m_0 }
                                        { \sqrt{ m_0^2 + \Lambda^2\, } - m_0 } \right] ,
\eeq
while $m_1,m_2$ are given in (\ref{61}) with $N=2$, namely, $m_1=-m_0$, $m_2=m_0$.

\vspace{1mm}

The weak coupling spectrum of the CP(1) model includes the tower of dyonic kinks
\beq
M_n=\left| \, m_D^{\rm CP(1)}+im_1 +in(m_2-m_1)\, \right|,
\label{weakCP1}
\eeq
where $n$ is an integer. The two states of the strong coupling spectrum (\ref{strongCP1}) belong 
to this tower with $n=0,1$. Other states from this tower as well as elementary (i.e. non-kink) states 
decay on the closed single CMS around the origin
in the $m^2_0$ complex plane \cite{SVZw}.

Now let us briefly discuss a more contrived situation, the CP(2) model with the $Z_3$ twisted masses. For simplicity we will restrict ourselves
to kinks interpolating between the third and first vacua  
($\sqrt{2}\sigma_3\approx m_3$ and  $\sqrt{2}\sigma_1\approx m_1$
in the large mass limit).
The strong coupling spectrum of the CP(2) model consists of three states
\beq
M_k^{13}=\left|\, m_D^{\rm CP(2)}+im_k\,\right|, \qquad k=1,2,3,
\label{strongCP2}
\eeq
where
\beq
\label{unod2}
	m_D^{\rm CP(2)} = -\, \frac{1}{2\pi} 
\left(  e^{2\pi i / 3} - 1 \right) 
	\bigg\{\, 3\sqrt[3]{ m_0^3 + \Lambda^3 } +
		\sum_j\, m_j\, \log  \frac{ \sqrt[3] { m_0^3 + \Lambda^3 } - m_j } 
{ \Lambda} \,\bigg\}\,
\eeq
and  the mass terms $m_k$ are given in (\ref{61}) with $N=3$.

\begin{figure}
\begin{center}
\epsfxsize=8.0cm
\epsfbox{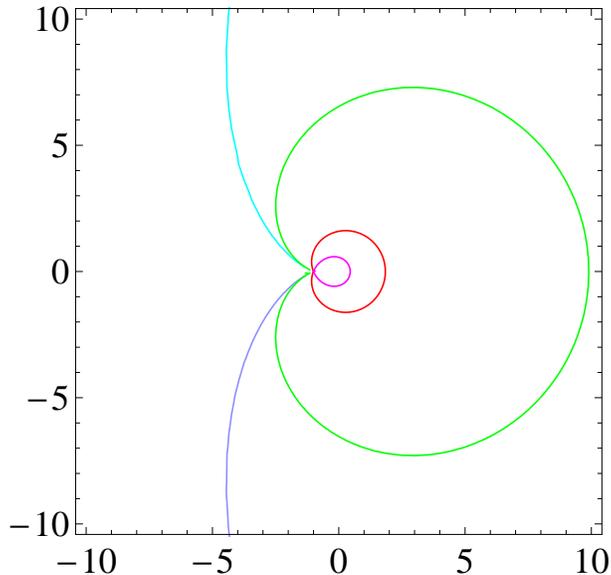}
\caption{\small The decay curves of CP$^2$ in $ m_0^3 $ plane. 
		}
\label{figCMS}
\end{center}
\end{figure}

The weak coupling spectrum of the CP(2) model includes two towers of dyonic kinks
\beqn
&& M_{n_1}^{13}=\left|\, m_D^{\rm CP(2)}+im_3 +in_1(m_1-m_3)\,\right|  \,\,\, {\rm and}
\nonumber\\[2mm] 
&&M_{n_2}^{13}=\left|\,m_D^{\rm CP(2)}+im_3 +in_2(m_1-m_3) +i(m_3-m_2)\,\right|,
\label{weakCP2}
\eeqn
plus elementary states.
Here $n_1$ and $n_2$ are integers. The two states of the strong coupling spectrum (\ref{strongCP2})
with $k=3,1$ both belong to the first tower. The
CP(2) model with the $Z_3$ masses has several CMS where all other states
except these two decay \cite{BSYcms}, see Fig~\ref{figCMS}. The third kink of the strong coupling
spectrum (with $k=2$) does not make it to the weak coupling domain. It decays on the most inner curve in
Fig.~\ref{figCMS}.

\section{Weighted CP\boldmath{$(N,M)$} models and \boldmath{$zn$} model}
\label{wei}

Considering ${\mathcal N}=2$ bulk theories of the type discussed in Sec. \ref{sec2} with $N_f>N$
(here $N_f$ is the number of flavors) we arrive at the so-called semilocal non-Abelian strings 
\cite{HT1,SYsem,Jsem,SYV}. Instead of $N_f$ we can introduce a positive number $M$,
\beq
N_f = N+M\,.
\eeq
The semilocal string solutions on the Higgs branches (typical
for multiflavor theories) usually are not fixed-radius strings, but, rather,
possess radial moduli $\rho^k$, also known as the size moduli (see   \cite{AchVas} for a  review of 
the Abelian semilocal strings). 

As previously, the 
orientational
moduli of the semilocal non-Abelian string can be described by a complex 
vector $n^P$ (here $P=1, ..., N$),
 while its size moduli are parametrized by a complex vector
$\rho^K$ (here $K=N+1, ..., N+M$). 

Originally it was conjectured \cite{HT1} (on the basis of string theory arguments) that 
the effective two-dimensional sigma model
describing low-energy dynamics on the semilocal string 
is the so-called weighted  CP$(N,M)$ model. This turned out to be not quite correct.
The world-sheet theory of the
moduli fields 
was derived in \cite{SYV,KSVY} and is known as the $zn$ model.
Its Lagrangian is
\beqn
{\mathcal L}_{zn} 
&=&
\left|\partial_{\alpha}(n^P \rho^K)\right|^2
 +\left|{\mathcal D}_{\alpha} n^{P}\right|^2 
 + \left|m_K -  m_P\right|^2 \left|n^{P}\right|^2 \left|\rho^{K}\right|^2 
\nonumber\\[3mm]
&+&
 \left|\sqrt{2}\sigma - m_{P}\right|^2\left|n^P\right|^2
+ iD  \left(|n^{P}|^2  -2\beta\right)^2
\,,
\nonumber\\[4mm] 
P&=&1,..., N\,,\qquad K=N+1,..., N + M\,,
\label{wcpp}
\eeqn
The $zn$ model so far remains largely unexplored. We refer the reader to the original papers for a brief
discussion.

The $zn$ model is similar but not identical to the weighted  CP$(N,M)$ model.
However, it was demonstrated \cite{SYV,KSVY} that its vacuum structure and  BPS spectrum
coincide with those  of the 
 ${\mathcal N}=2$ weighted CP$(N,M)$ model. Moreover, at $N\to \infty$ the $zn$ model
 and the weighted CP$(N,M)$ model coincide. Thus, the difference between them lies in the non-BPS
 sector at finite $N$.
 
 Technically, it seems more convenient  to work with the weighted  CP$(N,M)$ model.
  The bosonic 
part of its Lagrangian 
is 
\beqn
{\mathcal L}_{\rm WCP} 
&=&
 \left|{\mathcal D}_{\alpha} n^{P}\right|^2 
 +\left|\tilde{\mathcal D}_{\alpha} \rho^K\right|^2+
 \left|\sqrt{2}\sigma-  m_P\right|^2 \left|n^{P}\right|^2 
\nonumber\\[3mm]
&+&
 \left|\sqrt{2}\sigma - m_{K}\right|^2\left|\rho^K\right|^2
+ iD  \left(|n^{P}|^2-|\rho^K|^2 -2\beta\right)^2
\,,
\nonumber\\[4mm] 
P&=&1,..., N\,,\qquad K=N+1,..., N + M\,,
\label{wcp}
\eeqn
where
\beq
\rule{0mm}{6mm}
{\mathcal D}_{\alpha}=\partial_{\alpha}-
iA_{\alpha}\,,\qquad \tilde{\mathcal D}_{\alpha}=\partial_{\alpha} =
iA_{\alpha}\,.
\eeq
The mass terms $M_K$ and $M_P$ in (\ref{wcp}) are viewed as generic in this section.

\vspace{2mm}

The fields $n^{P}$ and $\rho^K$ have the opposite 
charges,  $+1$ and $-1$, with respect to the auxiliary U(1) gauge field.
This seemingly insignificant detail is crucial. Strictly speaking, the name `weighted CP' model is misleading
since the geometry of the target-space following from (\ref{wcp}) has nothing to do with the CP$(N-1)$ geometry  
in which all target space covariant quantities reduce to the metric, see (\ref{16}). The weighted CP models are not even renormalizable in the usual sense of this word. Nevertheless, the large-$N$ solution exists and is unique \cite{Koroteev1}.
We will discuss it in more detail in Sec. \ref{KMV}.


\section{Heterotic models}
\label{hetmo}

Heterotic two-dimensional models we will discuss below have two chiral supercharges, say, $Q_L$ and $\bar{Q}_L$
with the defining anticommutator
\beq
\{Q_L,\, \bar{Q}_L\} = 2 (H-P)\,.
\label{ai1}
\eeq
They are known as ${\mathcal N}= (0,2)$ supersymmetric sigma models.\footnote{In Sec. \ref{KMV} we will briefly comment on a ${\mathcal N}= (0,1)$ model.}
Previously they were studied mainly from the mathematical perspective \cite{west,Witten:2005px,bai3,Jia}. They can be divided into two classes: the so-called minimal and nonminimal models. 
This classification is in a sense similar to pure Yang-Mills theories and Yang-Mills theories with matter.
Later we will explain the difference between these two classes in more detail. In particular,
the minimal CP(1) model was considered in \cite{bai2}. This minimal model cannot be extended
to CP$(N-1)$ with $N>2$. The general hypercurrent structure in ${\mathcal N}= (0,2)$
was analyzed in \cite{bai4}.
In  what follows we will focus on those heterotic two-dimensional models
that are obtained on the world sheet of non-Abelian strings.

\subsection{How heterotic models appear}
\label{hhma}

If the bulk four-dimensional theory has ${\mathcal N}=2$ and supports 1/2-BPS strings,
then the low-energy theory on its world sheet  has four supercharges and, thus, possesses ${\mathcal N}= (2,2)$
supersymmetry. Now, if we slightly deform the bulk theory breaking ${\mathcal N}=2$ down to ${\mathcal N}=1$
we will have four supercharges in the bulk. For small deformations BPS saturation remains intact and so does the the target space of the two-dimensional sigma model. Now, the world-sheet model must have two, not four supercharges.
However, Zumino's theorem tells us that given a K\"ahler target space any supersymmetric nonchiral model is automatically uplifted to ${\mathcal N}= (2,2)$, i.e. four supercharges.

A way out was suggested by Edalati and Tong \cite{Edalati:2007vk} who conjectured a nonminimal ${\mathcal N}= (0,2)$ model on the string world sheet in the case of nonvanishing ${\mathcal N}=2$ breaking deformation in the bulk (see also
\cite{bai5}). This nonminimal theory was derived by Shifman and Yung \cite{Shifman:2008wv} 
from the analysis of the string solution.
The nonminimal theory, as it emerged  on the string world sheet, has no twisted masses.
In fact, even today we do not know which bulk theory might result in the nonminimal heterotic model with twisted masses. However, the inclusion of the twisted masses is straightforward in the two-dimensional model {\em per se},
without any reference to the bulk theory. This is the model to be discussed below too.

 Large-$N$ solutions of the heterotic models generically exhibit spontaneous breaking of supersymmetry. For 
 nonvanishing twisted masses
 this breaking occurs at the tree level. 

\subsection{Minimal vs. nonminimal CP\boldmath{$(N-1)$} models with \boldmath{${\mathcal N}= (0,2)$}
supersymmetry}
\label{mnvsnmn}

\subsubsection{Geometric formulation}
\label{geformu}

The minimal model can be obtained from (\ref{skinetic}) by keeping only left-handed fermions and discarding
all right-handed fermions, 
\beq
{\cal L}_{{\mathcal N}=(0,2)} 
= G_{i\bar j} \left[\rule{0mm}{6mm}\partial^\mu \phi^{\dagger\,\bar j}\, \partial_\mu\phi^{i}
+i\bar \psi^{\bar j}_L  {\mathcal D}_{R}\psi^{i}_L\right],
\label{ai2}
\eeq
where 
\beq
{\mathcal D}_{R}\psi_L^i = \partial_R\psi_L^i +\Gamma^i_{kl}(\partial_R  \phi^k)\psi^l_L\,,
\label{ai3}
\eeq
and
\beq
\partial_R\equiv \partial_t -\partial_z\,.
\label{ai4}
\eeq
The fields $\phi$ and $\psi_L$ form an $(0,2)$ supermultiplet.
In terms of ${\mathcal N}= (0,2)$ superfields \cite{west} one can act as follows.
Introduce a superfield
\beq
A=\phi(x_{R}+2i\theta^\dagger\theta,\,x_{L})+\sqrt{2}\,\theta\,\psi_L(x_{R}+2i\theta^\dagger\theta,\, x_{L})\,,
\label{ai5}
\eeq
where $\theta$ is a single (right-handed) complex Grassmann variable on the 
 $(0,2)$ superspace, and
\beq
x_L = t-z\equiv x^0 - x^1\,,\qquad x_R = t+z \equiv x^0 + x^1\,.
\eeq
Then
\beqn
\mathcal{L}_{\rm min} &= &\frac{1}{2}  \int   {d \theta d \bar\theta} \left[
\rule{0mm}{5mm}
K_i(A, A^\dagger) i{\partial_{R}} A^i + {\rm H.c.}\right]
\nonumber\\[2mm]
&=&-\frac{1}{4}\int {d\theta}\, G_{i\bar j}(A, A^\dagger) ({\bar{D}A^{\dagger \bar j})
\,
i\partial_{R}} A^i 
+ {\rm H.c.}
\label{ai7}
\eeqn

\vspace{2mm}

\noindent
{\em Warning:  Due to an anomaly pointed out in \cite{Moore} the heterotic minimal model is self-consistent only for CP(1) (see also \cite{C4}). Minimal CP$(N-1)$ models with $N>2$ do not exist. However, minimal heterotic O$(N)$ models exist for any $N$. For $N>3$ they have $(0,1)$ supersymmetry. For $ N=3$ we have ${O}(3) = {CP}(1)$. Nonminimal models presented in (\ref{ai8}) exist for CP$(N-1)$ at any $N$.}

\vspace{2mm}

Alternatively, one can start from Eq. (\ref{22}) and discard all terms containing $\xi_R$.

One last remark is in order here concerning the minimal CP(1) model presented in (\ref{ai7}). This is a strongly coupled theory. Since large-$N$ expansion is unavailable, we cannot solve it  by virtue of the large-$N$ expansion (we will apply it, however, to nonminimal 
heterotic CP$(N-1)$). Nevertheless, one feature of this model is known. As was shown in \cite{bai4}, current algebra in this model allows for a nonperturbative Schwinger term (see Eq. (5.7) in \cite{bai4}), namely, $C\sim \Lambda^2
\sim M_{\rm uv}^2 \exp\left(-\frac{4\pi}{g_0^2}
\right)$.
This Schwinger term is saturated by a single instanton due to the fact that in the model at hand it has 
just two fermion zero modes. The occurrence of this Swinger term implies spontaneous supersymmetry breaking. The interpolating field for Goldstino is
\beq
g\sim R_{i\bar j} \left( \partial_R \phi^i\right) \bar \psi_L^{\bar j}\,. 
\eeq
Spontaneous breaking of supersymmetry will be explicit in the large-$N$ solution of  the nonminimal 
heterotic CP$(N-1)$.

\vspace{5mm}

The bulk theories supporting non-Abelian strings are usually obtained by deforming  ${\mathcal N}=2$ theories  by a mass term of the adjoint superfield which breaks bulk supersymmetry down to ${\mathcal N}=1$. In this case the moduli fields on the string include all those inherent to the ${\mathcal N}= (2,2)$ CP$(N-1)$ model
plus an extra ${\mathcal N}= (0,2)$ supermultiplet with a peculiar interaction.
The heterotic model obtained in this way is to be referred to as
nonminimal. In the geometric formulation its Lagrangian is
\beqn
&&\hspace{-10mm}\mathcal{L} =
G_{i\bar j}\left[\partial_{R}\phi^{\dagger \bar j} \partial_{L}\phi^{i}+\psi_{L}^{\dagger \bar j}\,i{\mathcal D}_{\!R}\,\psi_{L}^{i}
+Z \psi_{R}^{\dagger \bar j}\,i{\mathcal D}_{\!L}\psi_{R}^{i}\right]
 + ZR_{i{\bar j} k {\bar l}}\,\psi_{L}^{\dagger \bar j}\psi_{L}^{i} \,\psi_{R}^{\dagger \bar l}\psi_{R}^{k} 
 \nonumber\\[2mm]
&&\hspace{-6mm}  +{\mathcal Z}\zeta_R^\dagger \, i\partial_L \, \zeta_R  +\! 
\left[\kappa\, \zeta_R  \,G_{i\bar j}\big( i\,\partial_{L}\phi^{\dagger \bar j}\big)\psi_R^{i}
+{\rm H.c.}\right] \! +\frac{{|\kappa |^2}}{Z} \zeta_R^\dagger\, \zeta_R
\big(G_{i\bar j}\,  \psi_L^{\dagger \bar j}\psi_L^{i}\big)\nonumber\\[2mm]
&&\hspace{-6mm}-\frac{|\kappa |^2}{\mathcal Z}
 \big(G_{i\bar j}\psi_{L}^{\dagger \bar j}\psi_{R}^{i}\big) \big(G_{k\bar l}\psi_{R}^{\dagger \bar l}\psi_{L}^{k}\big)\,.
 \label{components}
 \label{ai8}
\eeqn
Here ${\mathcal D}_{\!L,R}$ are covariant derivatives, 
\beq
{\mathcal D}_{\!L,R}\,\psi_{R,L}^{i}=\partial_{L,R}\,\psi_{R,L}^{i}+\Gamma^{i}_{kl}\,\partial_{L,R}\,\phi^{k}\,\psi_{R,L}^{l}\,. 
\label{ai9}
\eeq
The first line in (\ref{ai8}) coincides with the (2,2) Lagrangian in Eq. (\ref{skinetic}). The second and third lines
present a heterotic deformation. The right-handed fermion field $\zeta_R$ is absent in the (2,2) model. 

In terms of superfields the nonminimal heterotic model
can be written as follows:
\beqn
\mathcal{L} 
\!&=&\!  -\frac{1}{2} \int \!  d \theta \left[\frac{1}{2}\, G_{i\bar j}(A, A^\dagger) ({\bar{D}A^{\dagger \bar j})
\,
i\partial_{R}} A^i 
-\kappa \,G_{i\bar j}(A, A^\dagger)  
(\bar{D}A^{\dagger \bar j}){\mathcal B}\,B^i+{\rm H.c.}\right]
\nonumber\\[3mm]
&~&+
\frac{1}{2}\!\int d^2 \theta\left[
Z\, G_{i\bar j}(A, A^\dagger)\, B^{\dagger\bar j} B^i  + {\mathcal Z} \mathcal{B}^\dagger \mathcal{B}\right],
\label{ai10}
\eeqn
where $\kappa$ is the deformation parameter, and the extra (compared to the minimal model) (0,2) superfields are
\beqn
&&{\mathcal B}=\zeta_R(x_{R}+2i\theta^\dagger\theta,\,x_{L})+\sqrt{2}\,\theta\,F_\zeta(x_{R}+2i\theta^\dagger\theta,\, x_{L})\,\,\,{\rm and}\nonumber\\[3mm]
&&B=\psi_R(x_{R}+2i\theta^\dagger\theta,\, x_{L})+\sqrt{2}\,\theta F_\psi(x_{R}+2i\theta^\dagger\theta,\, x_{L})\,.
\label{ai11}
\eeqn
On mass shell both ${\mathcal B}$ and $B$ contain one fermion degree of freedom, $\zeta_R$ and $\psi_R$, respectively.

\subsubsection{Gauged formulation} 
\label{gaugeform}

The gauged formulation is most convenient for large-$N$ solution. Translation of the Lagrangian 
(\ref{ai8}) in the gauged formulation yields the Lagrangian (\ref{160}) presented in the Appendix. Technically, it is slightly more
convenient to work with an equivalent Lagrangian 
\beqn
{\mathcal L}_{(0,2)}
&=& \!\!
 \bar{\zeta}_R \, i
\partial_L
\, \zeta_R
+  
\left[ {2}i\,\omega\,\bar{\lambda}_{L}\,
\zeta_R  + {\rm H.c.}\right]
\nonumber\\[3mm]
&+& \!\!
 |{\mathcal D}_{\mu} n^{l}|^2 +2|\sigma|^2 |n^{l}|^2 + iD \left(|n^{l}|^2-2\beta\right)
\nonumber\\[3mm]
 &+&\!\!
 \bar{\xi}_{lR}\,i {\mathcal D}_{L}\, \xi^{l}_R
+ \bar{\xi}_{lL}\,i{\mathcal D}_{R}\, \xi^{l}_L
\nonumber\\[3mm]
 &+& \!\!
\left[ i\sqrt{2}\,\sigma\,\bar{\xi}_{lR}\xi^l_L
+
 i\sqrt{2}\,\bar{n}_l\,(\lambda_R\xi^l_L-
\lambda_L\xi^l_R) + {\rm H.c.} \right]
+\!\!
4\,|\omega|^2\left|\sigma
\right|^2  \nonumber\\
[2mm]
\label{cpg02p}
\eeqn
(equation (\ref{cpg02}) in the Appendix). The proof of equivalence is outlined in the Appendix.
The deformation constant $\omega$ is related to $\kappa$ in (\ref{ai8}) as follows:
\beqn
\kappa\beta &=& \frac{\omega}{\sqrt 2}\,, 
\label{aaaaa}
\\[2mm]
{\mathcal Z}_0 &=&  1+\frac{|\omega |^2}{\beta}\,,\qquad Z=1\,.\label{114}
\eeqn
The deformation parameter ${\omega}$ is renormalization-group invariant, see Sec. \ref{befu}.
In the large-$N$ solution we will see that physical effects are determined by an $N$-independent 
deformation parameter,
\beq
u=\frac{8\pi}{N} |\omega |^2 = \frac{16\pi}{Ng^2} \frac{\kappa^2}{g^2}\,. 
\label{116p}
\eeq
\vspace{2mm}
Both constants, $\kappa^2$ and $g^2$ 
scale with $N$ as $1/N$.

\subsubsection{Twisted masses}
\label{twm}

Twisted masses were added in \cite{BSY,L2}. The corresponding expressions are quite bulky. The interested reader is referred to the original publications. A novel element worth noting is as follows. In the absence of the heterotic deformation the CP$(N-1)$ model has $N-1$ complex twisted mass parameters, see Sec. \ref{twima}. With $\kappa
\neq 0$ the number of independent complex mass parameters generally speaking increases. In the generic case 
the nonminimal (0,2) model will have $N$ independent mass parameters. 

\subsection{Beta functions}
\label{befu}

All models under consideration in this review paper are asymptotically free. As was noted by Polyakov in 1975 \cite{poly}
at one loop only the bosonic fields contribute to the $\beta$ functions. Fermion contribution shows up at the 
two-loop level.

The exact all-loop $\beta$ function in the minimal CP(1) model was found in \cite{C3}.
It has the form
\beq
\beta_{g\,{\rm (0,2)\, min}} = -\frac{g^4}{2\pi} \left(1-\frac{g^2}{4\pi} 
\right)^{-1}\,,
\label{ai13}
\eeq
where $g^2$ is the coupling constant in the K\"ahler potential and metric.

Its structure is perfectly analogous to that of the NSVZ $\beta$ function in four-dimensional
${\mathcal N}=1$ gluodynamics \cite{NSVZ,SVa}. In fact, the above two-dimensional model is the closest analog of ${\mathcal N}=1$ four-dimensional theories one can think of. One can show 
\cite{C3} that the analogy extends further than Eq. (\ref{ai13}) and is maintained when one introduces ``matter" fields.
Then the $\beta$ function (\ref{ai13}) acquires a numerator typical of the NSVZ $\beta$ function in the presence
of matter. 

In the nonminimal model one deals with two coupling constants, $g^2$ appearing in the metric, and the deformation parameter $\kappa$. At one loop the corresponding $\beta$ functions were calculated in \cite{C2}, while the two-loop loop corrections and an exact relation 
between $\beta_g$ and the anomalous dimensions $\gamma$ were found in \cite{C4} (see also \cite{C2}),
\beqn
\beta_{g}\!=\!\mu\,\frac{dg^{2}}{d\mu}\!=\!-\frac{g^{2}}{4\pi}\,\frac{T_{G}\,g^{2}\left(1+\gamma_{\psi_R}/2
\right) - {h}^{2}\left(\gamma_{\psi_{R}}+\gamma_{\zeta}\right)}{1-({h}^{2}/4\pi)}
\label{11}
\label{betagh}
\eeqn
where $\gamma_{\zeta}$ and $\gamma_{\psi_R}$ are the anomalous dimensions of the
corresponding fields, which to the leading order are proportional to
\beq
h^{2}=\frac{|\kappa|^{2}}{Z{\mathcal Z}}\,,
\label{ai15}
\eeq
Here $Z$ and ${\mathcal Z}$ are filed renormalization constants for $\psi_R$ and $\zeta_R$ respectively (for their definition see \cite{C4}). 
At one  loop \cite{C2}
\beq
 \gamma\equiv \gamma_{\psi_{R}}+\gamma_{\zeta}= \frac{N\,{h}^{2}}{2\pi}\,.
 \label{ai16}
\eeq
The two-loop anomalous dimensions (which are also known \cite{C4}) give us explicit expression for $\beta_g$ at three loops. 

\begin{figure}[h]
\epsfxsize=8cm
\centerline{\epsfbox{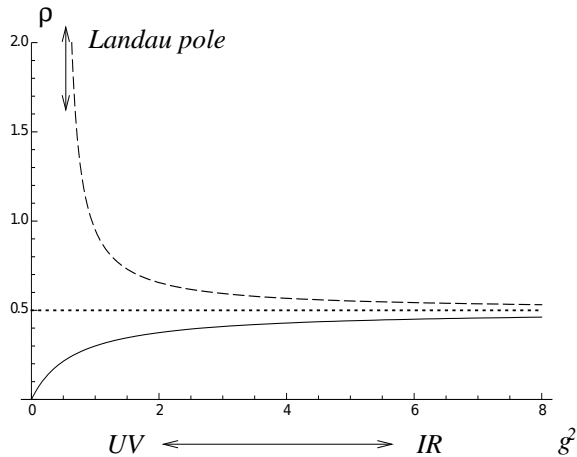}}
\caption{\small
Infrared fixed point in $\beta_\rho$.}
\label{abc}
\end{figure}

One can view $h^2$ as the second coupling constant -- the one responsible for the ${\mathcal N} =(2,2) \to (0,2)$ breaking.
It is convenient to consider the ratio
\beq
\rho \equiv h^2/g^2\,.
\eeq
the exact relation for the corresponding $\beta$ function is
\beq
\beta_\rho = \rho\left[\frac{1}{g^2}\,\beta_g +\gamma
\right].
\eeq
An explicit expression for $\beta_\rho$ exists \cite{C1,C4} up to two loops,
\beq
\beta^{(2)}_{\rho}=N\,\frac{g^{2}}{2\pi}\,\frac{\rho}{1-(h^{2}/4\pi)}\left(\rho -\frac{1}{2}\right).
\eeq
It has an infrared fixed pint at $\rho=1/2$ (see Fig. \ref{abc}). 
Whether the fixed point at $\rho =1/2$ is an exact statement or it does not hold in the third and higher loops is not known.

Another consequence from Eqs. (\ref{betagh}) and (\ref{ai16}) is as follows. In the limit $N\to \infty$ the constant
$h^2$ scales as $1/N$, implying that $\beta_{g}$ reduces to one loop and becomes exactly the same as in the undeformed ${\mathcal N}=(2,2)$ CP$(N-1)$ model. This is in full agreement with the large-$N$ solution of the nonminimal heterotic model to be presented below. The combination ${\kappa^2}/{g^4}$
is renormalization-group invariant,
\beq
\frac{\kappa^2}{g^4} = {\rm RGI}\,,
\eeq
cf. Eqs. (\ref{aaaaa}) and(\ref{116p}).

\subsection{Large-\boldmath{$N$} solution of nonminimal CP\boldmath{$(N-1)$}}
\label{lnsncp}

This model was solved with the $Z_N$ symmetric twisted masses \cite{L2} and arbitrary value of the mass parameter $m_0$. This solution includes of course
the massless heterotic model \cite{L1} as a limiting case $m_0=0$. Therefore, we will pass directly to the nonminimal model with the $Z_N$ symmetric twisted masses.

One brief remark is in order before this passage.
At small values of $u$, vanishing mass parameter $m_0$,  and {\em arbitrary} (i.e. not necessarily large)  
$N$ it is easy to find both the Goldstino
and the vacuum energy,
\beq
g\sim \omega\,  \left\langle R_{i\bar j}\, \bar\psi_R^{\bar j}\psi_L^i \right\rangle_{\rm vac} \zeta_R\,,
\eeq
where the vacuum averaging is performed in the undeformed $(2,2)$ massless CP$(N-1)$ model, see Eqs. 
(6.26) and (6.27) in \cite{Shifman:2008wv}. The extra right-handed field $\zeta_R$ plays the role of Goldstino.

\begin{figure}
\epsfxsize=11cm
\centerline{\includegraphics[width=11.5cm,keepaspectratio]{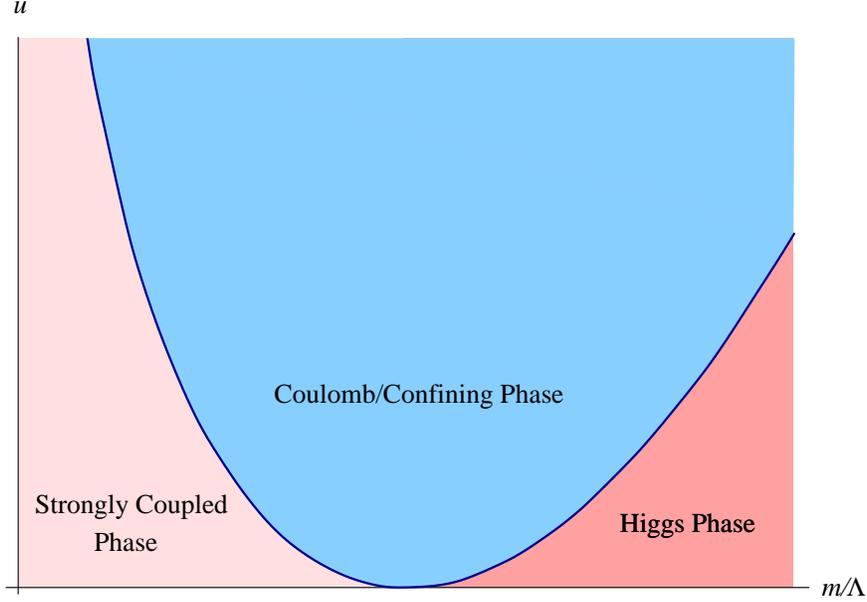}}
\caption{\small 
The phase diagram of the twisted-mass deformed heterotic CP($N-1$) theory
in the plane $u$ and $m_0$ where $m_0$ is assumed to be real.
The parameter $u$ denotes the amount of deformation, $ u = \frac{8\pi}{N} |\omega|^2 $.
}
\label{fig:higgsborder}
\end{figure}

Now, let us diñcuss the solution found in  \cite{L2}. Conceptually, the strategy of solving this model at 
large $N$ is similar to that described in Sec. \ref{lnswtm}.
Since in the model at hand
we have two parameters, $u$ and $m_0$, we discover a rather rich and  not quite trivial  phase diagram,
in which we observe phases with broken or unbroken $Z_N$ symmetry. If $u\neq 0$ we have two phases with the broken $Z_N$ symmetry, on the left and on the right in Fig. \ref{fig:higgsborder}. 
The first $Z_N$ phase is strongly coupled, the second (the Higgs phase) is weakly coupled.
In the middle lies the phase of unbroken $Z_N$ symmetry, in which the vacuum is unique, the photon does not acquire a mass, and the corresponding dynamical regime is that of charge confinement. 

Analytical solution for the vacuum structure is easier to obtain at large deformations, $u\gg 1$.

\subsubsection{ Strong coupling phase with broken \boldmath{$Z_N$}}
\label{scpwbz}

This phase occurs at very small masses, namely,
\beq
m_0\le \Lambda\,e^{-u/2}\,,\qquad u\gg 1\,.
\label{scphmass}
\eeq
In this phase we have 
\beq
|n|=0,\qquad iD \approx  \Lambda^2 \,,
\label{scphn}
\eeq
while the vacuum value of the $\sigma$ field is
\beq
\sqrt{2}\left\langle \sigma\right\rangle_{\rm vac}=e^{\frac{2\pi i}{N}k}\;\Lambda\,e^{-u/2},\qquad k= 1, ... , (N)\, .
\label{scphsigma}
\eeq
The vacuum value of $\sigma$ is exponentially small at large $u$. The bound $m_0<|\sqrt{2}\sigma|$
translates into the condition (\ref{scphmass}) for  $m_0$. For simplicity we will assume in this section
$m_0$ to be real and positive.

We  have $N$ degenerate vacua in this phase. The  chiral $Z_{2N}$ symmetry is broken
down to $Z_2$, the order parameter is $\langle \sigma\rangle$. Moreover, the absolute 
value of $\sigma$ in these vacua does not depend on $m$.
This solution essentially coincides with one obtained in \cite{L1} in the massless case. In this aspect the situation
is quite similar to the strong coupling phase of  the ${\mathcal N}=(2,2)$ model.  
The difference is that the absolute value of $\sigma$ depends now on $u$ and becomes exponentially
small in the limit $u\gg 1$.

\begin{figure}
\epsfxsize=10cm
\centerline{\input{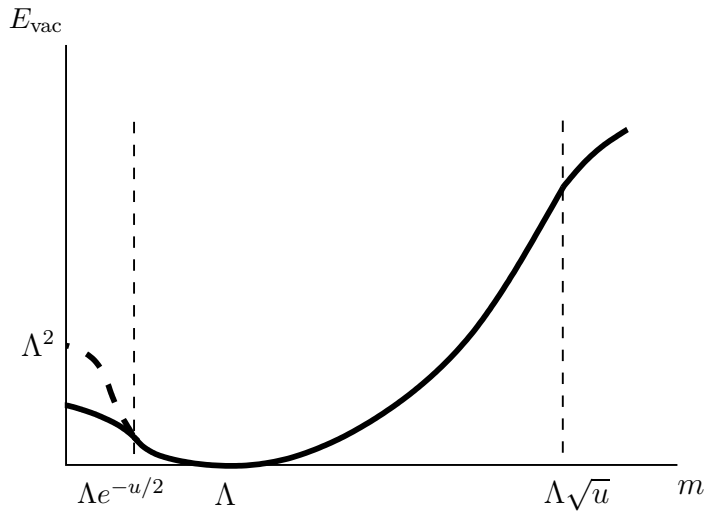}}
\caption{\small Vacuum energy density vs. $m_0$. The dashed line shows the behavior of the
energy density (\ref{confEvac}) extrapolated into the strong coupling region.} 
\label{figvacE}
\end{figure}

The vacuum energy is positive (see Fig. \ref{figvacE}).  Supersymmetry is spontaneously broken.

\subsubsection{Coulomb/confining phase}
\label{subscoulco}

Now we increase $m_0$ above the bound (\ref{scphmass}). The 
exponentially small $\sigma_{\rm vac}$ solution no longer exists. The only solution is
\beq
\langle \sigma\rangle_{\rm vac}=0\,.
\label{confsigma}
\eeq
In addition, Eq.~(\ref{scphn}) implies
\beq
|n|=0,\qquad iD= \Lambda^2-m^2\,.
\label{confnD}
\eeq
This solution describes a single $Z_N$ symmetric vacuum. All other vacua are lifted
and become quasivacua (metastable at large $N$). This phase is quite similar to the Coulomb/confining phase
of nonsupersymmetric CP$(N-1)$ model without twisted masses \cite{gfw}. The presence 
of small splittings between quasivacua produces a linear rising confining potential between kinks
that interpolate between, say, the true vacuum and the lowest quasivacuum \cite{odi},
see also the review \cite{SYbook}. Alternatively, this is a Coulomb interaction between charged particles due to a massless photon that results in confinement.

There is a phase transition (most likely  of the
second order) that separates these phases. As a rule, 
one does not 
have phase transitions in supersymmetric theories. However, in the model at hand  supersymmetry is
broken (in fact, it is broken already at the classical level \cite{BSY}); therefore, the emergence of 
a phase transition is not too surprising.  

One can calculate the vacuum energy  explicitly to see  the degree of supersymmetry breaking.
Substituting (\ref{confsigma}) and (\ref{confnD}) in the effective potential 
one gets
\beq
E_{\rm vac}^{\rm Coulomb}=\frac{N}{4\pi}\left[ \Lambda^2- m^2_0+ m_0^2\ln {\frac{m^2_0}{\Lambda^2}}\right].
\label{confEvac}
\eeq
see Fig.~\ref{figvacE}. At $m_0=\Lambda$ the vacuum energy vanishes
in the large-$N$ limit implying a supersymmetry restoration. Most likely, this vanishing will be lifted
by $1/N$ corrections, so that supersymmetry is always spontaneously broken.

\subsubsection{ Higgs phase}
\label{subshiggph}

The Higgs (weakly coupled) phase takes place in the model 
under consideration at large $m_0$,
\beq
m_0 > \sqrt{u}\Lambda\,,\qquad {\rm if} \,\,u \gg  1\,.
\label{Hphmass}
\eeq
In this phase $|n|$ develops a VEV, which is a clear-cut signal of the $Z_N$ symmetry breaking.
Thus, we conclude that
\beq
|n|^2_{\rm vac} 
=
\frac{N}{4\pi}\, \ln \frac{\sqrt{2}\sigma\, m_0}{\Lambda^2} 
\sim 
  \frac{N}{4\pi}\, \ln \frac{ m^2_0}{u\,\Lambda^2} 
\label{hujthree}
\eeq
in each of the $N$ vacua in the Higgs phase. 
were 
\beq
\sqrt{2}| \sigma |_{\rm vac} =  \left(\frac{8\pi}{N} \right)\,\frac{m_{{0}}}{u}\, ,
\label{Hphsigma}
\eeq

We have $N$ degenerate vacua again, as in the strongly coupled phase. In each of them $|\sigma |$ is small ($\sim m_0/u$) but
nonvanishing. The $Z_{N}$ chiral symmetry is  broken. Clearly, the 
Higgs phase is separated form the Coulomb/confining phase (where $Z_{N}$ is unbroken) by
a phase transition. 

\subsubsection{Goldstino}
\label{golds}

In this section we limit ourselves to the large-$N$ solution of the massless heterotic model (\ref{cpg02p}), 
derived from the bulk theory  in \cite{L1}.
Due to the spontaneous supersymmetry breaking we have a massless Goldstino fermion in the 
world-sheet theory.
To check this explicitly  one can  analyze the
one-loop effective Lagrangian calculated in \cite{L1}.
The appropriate fermionic part of the effective Lagrangian is
\beqn
{\mathcal L}_{\rm eff}^{\rm ferm}
&=&
\frac1{e^2_{\lambda}}\,\bar{\lambda}_{R}\,i\,\partial_{L}\,  \lambda_R
+\frac1{e^2_{\lambda}}\,\bar{\lambda}_{L}\,i\, \partial_{R}\, \lambda_L
+\frac12 \, \bar{\zeta}_R \, i
\, \partial_L \, \zeta_R
\nonumber\\[2mm]
 &+& 
\left[ \rule{0mm}{5mm}
i\sqrt{2}\,\Gamma\,\bar{\sigma}\,\bar{\lambda}_{L}\lambda_R  
+ \sqrt{2}\,i\,\omega\,\bar{\lambda}_{L}\,
\zeta_R  +{\rm H.c.}\right] \,,
\label{effferm}
\eeqn
where the one-loop couplings $e_{\lambda}$ and $\Gamma$ were calculated in \cite{L1}.

 First, we diagonalize the mass matrix for the $\zeta_R$, $\lambda_R$ and 
$\lambda_L$ fermions. Equating the determinant of this matrix to zero
produces the following equation for the mass eigenvalues $m$:
\beq
m^3-m\left( 2|\sigma|^2 \,\Gamma^2\,e_{\lambda}^4+4\,\omega^2\,e_{\lambda}^2\right)=0\,.
\label{detmass}
\eeq
For any $\omega$ we have a  vanishing eigenvalue corresponding to a massless 
Goldstino. Clearly, at small $\omega$ this fermion coincides 
with $\zeta_R$ (with an $O(\omega)$ admixture from the $\lambda$ fermions).

At large $u$ 
$$e_{\lambda}\sim \Lambda\,\,\,{\rm and}\,\,\, \Gamma \sim u/\Lambda^2\,,$$ while 
 $\sigma$ is given by  (\ref{scphsigma}). Thus, the last term in the second 
line in (\ref{effferm}) dominates, giving masses
to $\zeta_R$, and $\lambda_L$. The role of Goldstino is assumed by  the $\lambda_R$ fermion field.

\subsection{Large \boldmath{$N$} in nonminimal heterotic 
weighted \\CP\boldmath{$(N,M)$} model}
\label{LNM}

The unperturbed ${\mathcal N}=(2,2)$ model was discussed in Sec. \ref{wei}.
It is obtained on the world sheet of semilocal strings supported in the bulk ${\mathcal N}=2$
theories if $N_f>N$ \cite{HT1,SYsem,Jsem,SYV}. In this case there are two distinct types of the moduli fields,
$\rho$ and $n$, (scale and orientation moduli, respectively), and we arrive at the so-called $zn$ model on the world sheet. 
Hanany and Tong suggested \cite{HT1} the weighted CP$(N,M)$ for the same purpose. 
Later it was shown that these two models lead to identical predictions in the large-$N$ limit. 

If one introduces a $\mu\,{\rm Tr }{\mathcal A}^2$
deformation in the bulk theory breaking ${\mathcal N}=2$ down to ${\mathcal N}=1$
one arrives at a heterotically deformed model on the world sheet.  As far as we know, 
no explicit derivation of the deformation term in two dimensions starting from the deformed bulk theory
has ever been carried out. A conjecture that this deformation term is identical to that
emerging in  the $N_f=N$ case was formulated in \cite{Koroteev1}. Then the two-dimensional model obtained in this way 
was further generalized to include
twisted masses of two types, corresponding to two types of the moduli fields, namely, the scale and orientational moduli.
To reduce the number of adjustable parameters, it was assumed that the first set of the twisted masses is $Z_M$ symmetric, while the second is $Z_N$ symmetric (cf.  (\ref{61})). As a result, there are two mass parameters
$m_0$ and $\mu_0$ plus two dimensionless parameters
\beq
\alpha= \frac{M}{N}\,\,\, {\rm and}\,\,\,  u\,.
\eeq
The limit $N\to \infty$ was assumed. The large-$N$ analysis of the vacuum structure and the spectrum of the
model  is very similar to that discussed
in Sec.~\ref{lnsncp}. Under these conditions  the model was solved \cite{Koroteev1}
and a rich structure discovered on the phase diagrams, including two distinct Higgs phases and two distinct Coulomb phases and various patterns of the $Z_{N,M}$ breaking. 
An interesting phenomenon was observed on a two-dimensional subspace of  mass parameter space on which a discrete $Z_{N-M}$ symmetry is preserved. As was expected, supersymmetry is spontaneously broken  for generic values of adjustable parameters. However, 
on a special curve in the parameter space we have the same phenomenon as at $m_0 =\Lambda$ in Fig. \ref{figvacE}.
Supersymmetry seems to be
restored at $N\to \infty$. A new branch opens up for special values of $m_0$ and $\mu_0$.
In much the same way as in Sec. \ref{subscoulco} one can expect that the vacuum energy on this curve will be lifted
in a subleading order in $1/N$.

\subsection{Large \boldmath{$N$} in  heterotic 
O\boldmath{$(N)$} model}
\label{KMV}

To begin with, a few words about the minimal $(0,1)$ ${\rm O}(N)$ model will be in order.
Assuming $N\geq4$ it is easy to obtain this model by truncating the standard $(1,1)$ model \cite{eddieyoung}, for a review see \cite{nsvzobz}.
To this end we introduce the $(0,1)$ superfield
\beq
N^a = S^a(x)+   \theta_R \, \psi_L^a (x)\,,\qquad a=1,2, ..., N\,,
\label{133}
\eeq
with the following Lagrangian \cite{west} (plus the standard constraint)
\beq
{\mathcal L}_{(0,1)\,\,{\rm min}} = \frac{1}{2g^2}\int d\theta_R \,\left( D_L N^a\right) \left(i\,\partial_R N^a\right)\,,
\qquad N^aN^a -1 =0\,,
\label{134}
\eeq
where $\psi_L^a$ is a Weyl-Majorana field, $\partial_R =\partial_t-\partial_z$  as usual,
and $$D_L = \frac{\partial}{\partial\theta_R} - i\theta_R\partial_L\,.$$
The constraint in (\ref{134}) can be implemented by adding an appropriate Lagrange multiplier term
\beq
\Delta {\mathcal L}_{(0,1)\,\,{\rm min}} = \int d\theta_R \, X
\left( N^aN^a -1 \right)\,,
\label{135}
\eeq
where 
\beq
X = \frac{1}{2g^2}\left(-\lambda_R + \theta_R D\right)\,.
\label{136}
\eeq
Note that, in contradistinction with the CP$(N-1)$ case, the minimal O$(N)$ model exists at all $N$.
The large-$N$ solution of the model (\ref{134}) is constructed in  much the same way as that
for nonsupersymmetric O$(N)$ model \cite{nsvzobz}. Supersymmetry is spontaneously broken, the constraint 
$S^aS^a =1$ is lifted, all $S^a$ fields acquire a mass while the $\psi_L$ fields remain massless. The field $\chi_R$ acquires a kinetic term. 

\vspace{3mm}

To construct a nonminimal heterotic model we will follow the same line of reasoning as in Sec. 
\ref{geformu}. In fact, in the geometric formulation one can use the Lagrangian (\ref{ai10}) with the replacement
of the K\"ahler metric of CP$(N-1)$ 	
by a real metric of the $N$-dimensional sphere, and assuming that the parameter $\theta$
in the definition of the superfields (\ref{ai5}) and (\ref{ai11}) is real.

A slightly different formulation is more convenient for the large-$N$ analysis, however. In addition to the 
$(0,1)$
superfield (\ref{133}) let us introduce 
two right-handed ``matter" superfields (both with one physical degree of freedom),
\beq
{\mathcal B}=\zeta_R (x)+\theta_R\,F_\zeta(x)\,,\qquad
B^a=\psi_R^a(x)+ \theta_R F^a_\psi(x)\,.
\label{137}
\eeq
The Lagrangian of the model can be written as
\beqn
{\mathcal L}_{(0,1)} &=&
\int d\theta_R \left\{  \frac{1}{2g^2}\left[ 
\rule{0mm}{4mm}
 \left( D_L N^a\right)\left(i\,\partial_R N^a\right)
+\left(D_L\,B^a\right) B^a\right] +\frac{1}{2}\left(D_L\,{\mathcal B}\right) {\mathcal B}\right.
\nonumber\\[3mm]
&-&
\frac{\kappa}{g^2}\,\left( D_L N^a\right)B^a  {\mathcal B}
- \left.
 X
\left( N^aN^a -1 \right) - \tilde X \left( N^aB^a  \right)
\rule{0mm}{6mm}\right\}
\,,
\label{138}
\eeqn
where the last two terms implement the constraints $S^aS^a =1$ and $S^a\psi^a_{L,R}=0$ (plus the
standard  relation for $F_\psi$, see \cite{nsvzobz})
and $\tilde X$ is an auxiliary field analogous to (\ref{136}), namely, 
\beq
 \tilde X = \frac{1}{g^2}\left(\sigma+ \theta_R\lambda_L\right)\,.
 \eeq

In components (after eliminating the auxiliary fields  $F_{\zeta ,\,\psi}$ and a rescaling needed to make kinetic terms canonical) the Lagrangian takes the form \cite{Koroteev2}
\beqn
{\mathcal L}_{(0,1)} &=&
 \frac{1}{2} \partial_L S^a \partial_R S^a+ 
\frac{i}{2} \psi_L^a\partial_R\psi_L^a +\frac{i}{2} \psi_R ^a\partial_L\psi_R^a 
+\frac{i}{2} \zeta_R \partial_L\zeta_R
\nonumber\\[3mm]
&+&
\beta_L\psi_R^aS^a + \chi_R\psi_L^aS^a-\frac{1}{2} \left(D+\sigma^2\right) S^aS^a
+\frac{1}{2} \, \frac{D}{g^2}
\nonumber\\[3mm]
&+&
\sigma\psi_L^a\psi_R^a + \kappa\left(i \partial_L S^a\right)\psi_R^a \zeta_R  +\frac{1}{2}\kappa^2 \sigma^2\,.
\eeqn
It is not difficult to calculate the effective potential as
a function of $D$ and $\sigma$ \cite{Koroteev2},
\beq
V_{\rm eff} = \frac{N}{8\pi}\left[D\log\frac{\Lambda^2}{D+\sigma^2}+\sigma^2\log\frac{\sigma^2}{\sigma^2+D}+D+u\sigma^2\right]\,,
\eeq
where  
\beq
u=\frac{4\pi\kappa^2}{g^4 N}\,.
\eeq
Minimizing the potential with respect to $D$ and $\sigma$ one finds two distinct vacua of the theory
\beq
\sigma_0 = \pm\Lambda {e}^{-\frac{u}{2}}\,, \qquad
D = \Lambda^2 - \sigma^2\,,
\eeq
which present continuations of two distinct vacua inherent to the supersymmetric $(1,1)$ limit of the model.
The ensuing vacuum energy is
\beq
E_{\rm vac} = \frac{N}{8\pi}\Lambda^2\left(1-{e}^{-u}\right)\,.
\eeq
Any nonvanishing value of $u$ results in the spontaneous breaking of supersymmetry. The spectrum of the model and, in particular, the Goldstino composition can be readily found too.

\section{In the uncharted waters}
\label{Fstring}

In Sec. \ref{sec2} we outlined the simplest prototype bulk theory
supporting non-Abelian strings which, in turn, give rise to the observed wealth of two-dimensional sigma models 
in Secs. \ref{bamo} -- \ref{hetmo}. Extending the bulk theory one can expect do derive novel sigma models on the string world sheet. In this section we will briefly discuss an extended construction resulting in the $(0,2)$ two-dimensional model 
which has never been discussed previously. Moreover, its geometric formulation is not yet known. 

For brevity of presentation we will stick to $N=2$ and $N_f \geq 2$, referring the reader to the original papers 
\cite{SYfstr,BSYadj} for the case of generic $N$. Unlike Sec. \ref{sec2} we will switch off the Fayet-Iliopoulos $D$ term (i.e. $\xi=0$ in the last term in Eq. (\ref{pot})) but, instead, switch on the
mass term for the adjoint fields ${\mathcal A}$,
\beq
{\mathcal W}_{{\rm def}}=
  \mu\,{\rm Tr}\,\Phi^2, \qquad \Phi\equiv\frac12\, {\mathcal A} + T^a\, {\mathcal A}^a
\label{msuperpotbr}
\eeq
in addition to non-vanishing mass terms for the bulk (s)quark fields \cite{SYfstr,BSYadj}. 
The deformation (\ref{msuperpotbr}) breaks bulk supersymmetry down to ${\mathcal N}=1$, generally speaking.

This leads to the following modification
of the bulk potential (\ref{pot}). The last two $F$ terms in the second line in (\ref{pot}) 
responsible for the squark condensation are replaced by
\beq
 2g^2_2\left| \tilde{q}_A T^a q^A +\frac{\mu}{\sqrt{2}}\, a^a\right|^2+
\frac{g^2_1}{2}\left| \tilde{q}_A q^A +\frac{N}{\sqrt{2}}\mu a \right|^2\,.
\eeq
Since  VEVs of the adjoint fields $a$ and $a^a$ are determined by squark masses (cf. (\ref{avev})) this leads to
the breaking of the color-flavor symmetry. The quark VEVs are no longer degenerate. Instead of (\ref{qvev})
 the quark VEVs take the form
 \beqn
\langle q^{kA}\rangle &=& \langle\bar{\tilde{q}}^{kA}\rangle=\frac1{\sqrt{2}}\,
\left(
\begin{array}{ccccc}
\sqrt{\xi_1} & 0 & 0 & \ldots & 0\\
0  & \sqrt{\xi_2} & 0 & \ldots & 0\\
\end{array}
\right),
\nonumber\\[3mm]
k&=&1,\, 2\,,\qquad A=1,...,N_f\, .
\label{qvevN}
\eeqn
The parameters $\xi_{1,2}$ in (\ref{qvevN}) in the  quasiclassical  approximation are 
\beq
\xi_{1,2} \approx 2\;\mu \, m_{1,2}\,.
\label{xiclass}
\eeq
These parameters can be made large in the large-$m$ limit even if  $\mu$ is  small, to ensure that the 
bulk theory is at weak coupling.

The squark condensation leads to the string formation. If $m_1\neq m_2$
these strings  have nondegenerate tensions. The U$(2)$ gauge group is broken down to
U(1)$\times$U(1) by the quark mass difference. To the leading order in $\mu$ 
each U(1) gauge factor supports it own BPS string. The string
tensions of two strings under consideration are  \cite{SYfstr}
\beq
T_{1,2}=2\pi|\xi_{1,2}|\,.
\label{ten}
\eeq
If $|m_1-m_2|\ll |m_{1,2}|$  these two strings can still be promoted to non-Abelian strings
with a shallow potential in the world-sheet theory.
As was mentioned, now  \ntwot supersymmetry is broken down to
\ntwoo  even to the leading order in $\mu$. For the single-trace deformation  (\ref{msuperpotbr})
the bosonic part of the world-sheet theory becomes  \cite{SYfstr,BSYadj}
\beq
{\mathcal L} ={\mathcal L}_{(2,2)}+ V_{\rm def} (\sigma),
\label{Sfstring}
\eeq
where ${\mathcal L}_{(2,2)}$ is the Lagrangian of  the \ntwot supersymmetric model (\ref{wcp}) while 
the deformation potential is 
\beq
V_{\rm def} (\sigma)=4\sqrt{2}\pi\,|\mu\sigma|\,.
\label{Vdefr=N}
\eeq
The deformation (\ref{Vdefr=N}) respects only $(0,2)$ superalgebra.

This potential is radically different from the $|\sigma|^2$ potential in the heterotic deformation (\ref{cpg02p}). The latter
potential arises on the non-Abelian string in the massless bulk theory with the Fayet-Iliopoulos $D$-term 
deformed by the superpotential (\ref{msuperpotbr}).

The total scalar potential is given by the sum of the twisted mass potential in (\ref{wcp}) and deformation 
(\ref{Vdefr=N}). Its  minima correspond to tensions of 
two elementary non-Abelian strings,
\beq
V(\sigma_{1,2})_{\rm def} = T_{1,2}\,.
\label{Vten}
\eeq
To see that this is the case we note that at 
small $\mu$ the vacuum values of $\sigma$ are still determined by the squark masses
$\sqrt{2} \sigma_{1,2}\approx m_{1,2}$ in the quasiclassical approximation. Then (\ref{Vten}) 
follows\,\footnote{ This statement is valid beyond the quasiclassical approximation (to all orders
in  $\Lambda/m_{1,2}$). In this case the $\sigma $ VEVs  are determined \cite{SYfstr} by the roots of the equation (\ref{sigmaeq}).}
 from (\ref{xiclass}). 

If $m_1\neq m_2$ the minima are nondegenerate.  Only the lowest-lying vacuum is stable.
The stability of the lowest vacuum in two dimensions means that only 
the lightest  non-Abelian string is stable, the other one is metastable. Moreover, since
generically the string tensions do not vanish, \ntwoo super\-symmetry is broken spontaneously already at 
the classical level \cite{SYfstr}. 

To conclude this section let us mention that at the generic quark masses the deformation (\ref{msuperpotbr})
leads to the emergence of a whole set of isolated vacua in the bulk theory, the so-called $r$ vacua, $r\le N$.
In each $r$ vacuum $r$ quarks and $(N-r-1)$ monopoles condense. The vacuum in (\ref{qvevN}) correspond to the 
$r=N$
 vacuum, with the maximal number of condensed quarks. The simplest example of $r<N$ vacuum, namely,
an  $r=N-1$
vacuum (with $r=N-1$ condensed squarks and no monopoles) was considered in \cite{SYquantstr}. This vacuum  also 
supports non-Abelian strings. However,  in 
contradistinction with the $r=N$ vacuum,  the two-dimensional theory on the string world-sheet
receives in this case nonperturbative corrections from the bulk, through the bulk gaugino condensate.
 Nonperturbative bulk effects deforming  the theory on the string  world sheet 
were found in \cite{SYquantstr} by virtue of the method of resolvents suggested by Gaiotto, Gukov 
and Seiberg for surface defects \cite{GGS}.

 \section{Conclusions}
 \label{concl}
 
 Forty years ago A. Polyakov emphasized that asymptotically free two-dimen\-sional sigma models
 could present the best laboratory for the four-dimensional Yang-Mills theories. This prophecy came true 
 in various aspects -- even more than it was anticipated. First and foremost, a remarkable 2D-4D correspondence
 was detected in supersymmetric theories (see \cite{SYbook,SYquantstr} and references therein):
 the BPS spectrum of the sigma models on the string world sheet proves to be in one-to-one correspondence
 with that in the bulk four-dimensional theory. Moreover, diverse two-dimensional sigma models 
{\em per se} exhibit nontrivial dynamical features which, quite unexpectedly, proved to be in close parallel
 with some features of four-dimensional Yang-Mills. Novel models continue to appear in the limelight.
Today the task of their exploration is highly challenging. This path is fruitful.

 \section*{Acknowledgments}
 \addcontentsline{toc}{section}{Acknowledgments}

This work  is supported in part by DOE grant DE-FG02-94ER40823. 
The work of A.Y. was  supported 
by  FTPI, University of Minnesota, 
by RFBR Grant No. 13-02-00042a 
and by Russian State Grant for 
Scientific Schools RSGSS-657512010.2.

\section*{Appendix: Various representations of the \\ nonminimal heterotic model}

\addcontentsline{toc}{section}{Appendix: Various representations of the nonminimal \\ heterotic model}

\renewcommand{\theequation}{A.\arabic{equation}}
\setcounter{equation}{0}

A nonmnimal heterotic deformation of the CP$(N-1)$ model was suggested in
\cite{Edalati:2007vk} by adding a twisted superpotential (0,2) term in the gauge representation
(\ref{22}) following from (\ref{sqed}) in the limit $e^2\to \infty $. 
In the appropriate normalization in components the corresponding Lagrangian is
\beqn
{\mathcal L}
&=& \!\!
 \bar{\zeta}_R \, i
\partial_L
\, \zeta_R
+  
\left[ {2}i\,\omega\,\bar{\lambda}_{L}\,
\zeta_R  + {\rm H.c.}\right]
\nonumber\\[3mm]
&+& \!\!
 |{\mathcal D}_{\mu} n^{l}|^2 +2|\sigma|^2 |n^{l}|^2 + iD \left(|n^{l}|^2-2\beta\right)
\nonumber\\[3mm]
 &+&\!\!
 \bar{\xi}_{lR}\,i {\mathcal D}_{L}\, \xi^{l}_R
+ \bar{\xi}_{lL}\,i{\mathcal D}_{R}\, \xi^{l}_L
\nonumber\\[3mm]
 &+& \!\!
\left[ i\sqrt{2}\,\sigma\,\bar{\xi}_{lR}\xi^l_L
+
 i\sqrt{2}\,\bar{n}_l\,(\lambda_R\xi^l_L-
\lambda_L\xi^l_R) + {\rm H.c.} \right]
+\!\!
4\,|\omega|^2\left|\sigma
\right|^2  
.
\label{cpg02}
\eeqn
In this form it was used\,\footnote{Note a different normalization of the $\zeta$ kinetic term in Eq. (2.13)
in \cite{L1}.} in the large-$N$ solution of the model in \cite{L1}.
The constraint on $\bar n \xi_R$ ensuing from (\ref{cpg02}) is
\beq
\bar n \xi_R = \sqrt{2} \bar \omega \bar\zeta_R\,.
\label{156}
\eeq
One can pass to the standard form of this constraint $\bar n \xi_R =0$ inherent to (2,2) supersymmetry 
by shifting the $\bar\xi$ and $\xi$ fields,	
\beq
\xi=\xi' + \frac{\bar\omega}{\sqrt{2}\,\beta}\,n\bar\zeta_R\,,\qquad
\bar\xi=\bar\xi' + \frac{\omega}{\sqrt{2}\,\beta}\,\bar{n}\,\zeta_R\,.
\label{157}
\eeq
In terms of $\xi', \,\,\bar\xi'$ 
\beq
\bar{\xi}_{lR}\,i {\mathcal D}_{L}\, \xi^{l}_R\to (\bar{\xi}_{lR})'\,i {\mathcal D}_{L}\, (\xi^{l}_R)'\,,
\label{158}
\eeq
and instead of the first and second terms in the first line in (\ref{cpg02}) one obtains
\beqn
&&\bar{\zeta}_R \, i
\partial_L
\, \zeta_R
+  
\left[ {2}i\,\omega\,\bar{\lambda}_{L}\,
\zeta_R  + {\rm H.c.}\right]
\nonumber\\[2mm]
&& \to 
\left(1+\frac{|\omega |^2}{\beta}
\right) \left(\bar{\zeta}_R \, i
\partial_L
\, \zeta_R\right)
+ \left[\frac{\bar\omega}{\sqrt{2}\,\beta} (\bar{\xi}_{lR})'\bar\zeta_R \,i {\partial}_{L}n^l 
+ {\rm H.c.}\right]
\label{159}
\eeqn
Now we can omit primes, $\xi' \to \xi$ in the transformed Lagrangian ${\mathcal L}$ replacing
(\ref{cpg02}).
The constraint (\ref{156})
is traded for the trilinear term in the transformed Lagrangian, 
\beqn
{\mathcal L}_{\rm trilin}
&=& \!\!
\left(1+\frac{|\omega |^2}{\beta}
\right) \left(\bar{\zeta}_R \, i
\partial_L
\, \zeta_R\right) \bar{\zeta}_R \, i
\partial_L
\, \zeta_R
+  
\left[\frac{\bar\omega}{\sqrt{2}\,\beta} (\bar{\xi}_{lR})'\bar\zeta_R \,i {\partial}_{L}n^l 
+ {\rm H.c.}\right]
\nonumber\\[3mm]
&+& \!\!
 |{\mathcal D}_{\mu} n^{l}|^2 +2|\sigma|^2 |n^{l}|^2 + iD \left(|n^{l}|^2-2\beta\right)
\nonumber\\[3mm]
 &+&\!\!
 \bar{\xi}_{lR}\,i {\mathcal D}_{L}\, \xi^{l}_R
+ \bar{\xi}_{lL}\,i{\mathcal D}_{R}\, \xi^{l}_L
\nonumber\\[3mm]
 &+& \!\!
\left[ i\sqrt{2}\,\sigma\,\bar{\xi}_{lR}\xi^l_L
+
 i\sqrt{2}\,\bar{n}_l\,(\lambda_R\xi^l_L-
\lambda_L\xi^l_R) + {\rm H.c.} \right]
+\!\!
4\,|\omega|^2\left|\sigma
\right|^2  
.
\label{160}
\eeqn
Equation (\ref{160}), being rewritten in the geometric form, identically coincides with (\ref{ai8})
provided that 
\beq
\kappa\beta = \frac{\omega}{\sqrt 2}\,, \qquad {\mathcal Z} =  1+\frac{|\omega |^2}{\beta}\,,\qquad Z=1\,.
\label{161}
\eeq

\end{document}